\documentclass[12pt]{article}
\usepackage{amsmath}
\usepackage{graphicx,psfrag,epsf}
\usepackage{tabularx}
\usepackage{enumerate}
\usepackage{natbib}
\usepackage{url} 
\usepackage{authblk}
\usepackage{wrapfig}
\RequirePackage{amsthm,amsmath,amsfonts,amssymb}
\usepackage{amsmath}
\usepackage{amssymb}
\usepackage{tikz}
\usetikzlibrary{arrows}
\usepackage{float} 
\usepackage{booktabs}
\usepackage{algorithm}
\usepackage{algpseudocode}
\usepackage{bbm}
\usepackage{ragged2e}
\usepackage{subcaption}
\usepackage{adjustbox}
\newtheorem{prop}{Proposition}
\newcommand{\ijbold}{(\mathbf{i},\mathbf{j})}
\newcommand{\bi}{\begin{itemize}}
\newcommand{\ei}{\end{itemize}}

\usepackage[shortlabels]{enumitem}
\usepackage{amsmath}
\usepackage{arydshln}
\usepackage{graphicx}  
\usepackage{url}
\usepackage{breakurl}
\usepackage{makecell}

\usepackage{caption}
\DeclareMathOperator*{\argmax}{arg\,max}

\newcommand{\blind}{1}

\begin{document}

\def\spacingset#1{\renewcommand{\baselinestretch}%
{#1}\small\normalsize} \spacingset{1}


\if1\blind
{
  \title{\bf BayesFLo: Bayesian fault localization of complex software systems}
    \small
    \author[1,2]{Yi Ji}
    \author[2]{Simon Mak}
    \author[1]{Ryan Lekivetz}
    \author[1]{Joseph Morgan}
    \affil[1]{JMP Statistical Discovery LLC, SAS Institute Inc.}
    \affil[2]{Department of Statistical Science,
Duke University}
    \maketitle
} \fi

\if0\blind
{
  \bigskip
  \bigskip
  \bigskip
  \begin{center}
    {\LARGE\bf BayesFLo: Bayesian fault localization of complex software systems}
\end{center}
  \medskip
} \fi

\bigskip
\begin{abstract}
    Software testing is essential for the reliable development of complex software systems. A key step in software testing is fault localization, which uses test data to pinpoint failure-inducing combinations for further diagnosis. Existing fault localization methods have two key limitations: they (i) largely do not incorporate domain and/or structural knowledge from test engineers, and (ii) do not provide a probabilistic assessment of risk for potential root causes. Such methods can thus fail to confidently whittle down the combinatorial number of potential root causes in complex systems, resulting in prohibitively high debugging costs. To address this, we propose a novel Bayesian fault localization framework called BayesFLo, which leverages a flexible Bayesian model for identifying potential root causes with probabilistic uncertainties. Using a carefully-specified prior on root cause probabilities, BayesFLo permits the integration of domain and structural knowledge via the principles of combination hierarchy and heredity, which capture the expected structure of failure-inducing combinations. We then develop new algorithms for efficient computation of posterior root cause probabilities, leveraging recent tools from integer programming and graph representations. Finally, we demonstrate the effectiveness of BayesFLo over existing methods in two fault localization case studies, the first on the Traffic Alert and Collision Avoidance System for aircraft collision avoidance, and the second on the Vulnerable Road User protection tests for safe autonomous driving. 
\end{abstract}

\noindent%
\vfill

\newpage
\spacingset{1.45} 



\section{Introduction}
\label{sec:Intro}
Software testing -- the process of executing a program with the intent of finding errors \citep{myers2004art} -- is an essential step in the development of robust software applications. Such testing aims to reveal and subsequently fix as many bugs (i.e., faults)
as possible prior to the deployment of a software application, thus greatly reducing the likelihood of encountering failures for the end-user. This is crucial in an era where nearly all facets of daily life involve human interaction with software applications. 
There are, however, two fundamental challenges. First, each software test can be time-consuming to perform. This involves not only running the software system itself, which can be intensive for modern scientific applications \citep{ji2021graphical,ji2024conglomerate}, but also determining whether the software deviates from its expected behavior. The latter, known as the ``oracle problem'' \citep{barr2014oracle}, typically requires 
an independent assessment of numerical accuracy of software outputs \citep{lekivetz2021testing}, which can be very costly.
Second, the number of test cases required for thorough software examination can easily be overwhelming. As ``bugs [tend to] lurk in corners and congregate at boundaries'' \citep{beizer2003software}, software testing typically focuses on boundary values and the \textit{combinations} of inputs, which can grow rapidly. For practical applications, it is thus wholly infeasible to exhaustively test all input combinations \citep{kumar2019review}, which can easily require billions of test cases.

These two fundamental challenges open up exciting new statistical directions for this application of rising importance. Such directions can roughly be categorized into two topics. The first is a careful {design} of test cases with the joint goals of identifying failure settings and diagnosing their root causes. Statistically, this can be viewed as an \textit{experimental design} problem for software fault diagnosis. There has been notable work on this front. Early work on one-factor-at-a-time designs \citep{frey2003role} explored the problem of unit testing \citep{runeson2006survey,khorikov2020unit}, which focuses on testing small functional code units, e.g., a single function or chunk of code. However, such testing cannot identify bugs that congregate at the interaction of such units, which often arise in large software applications. Another approach is pairwise testing \citep{bach2004pairwise}, which examines all {pairwise} combinations of inputs; test case generation for this setting has been explored in \cite{tai2002test}. A more general approach is \textit{combinatorial testing} \citep{nie2011survey}, which investigates combinations involving more than two inputs. For combinatorial testing, the design of choice is a \textit{covering array} (CA; \citealp{colbourn2004combinatorial}), which aims to represent (or ``cover'') each combination of inputs (up to a specified order) at least once. CAs enable the detection of failures from limited test runs; see Section \ref{sec:mot} for further discussion.



With the initial test cases conducted and failures detected, the second direction is \textit{fault localization} \citep{wong2023software}: the use of this test data to pinpoint root causes. This is a highly challenging problem due to the overwhelming number of scenarios to consider for potential root causes. To see why, consider a software application with $I=10$ input factors each with two levels. For this system, there are a total of $\sum_{i=1}^{10}\binom{10}{i} 2^i=59,048$ input combinations that might be potential root causes; these consist of all single factors, 2-factor combinations, and so on. Since each combination is either a root cause or not, this results in $2^{59048}$ different scenarios to consider for potential root causes! Fault localization requires gauging which of these many scenarios are likely given test outcomes, which is a computationally intensive task \citep{wong2023software}, even for systems with a moderate number of inputs and few failed test cases. Furthermore, the debugging cost for investigating each such scenario can be considerably high \citep{boehm1984software,humphrey1995discipline}. For example, for critical flight software systems at the Jet Propulsion Laboratory \citep{bush1990improving}, each debugging activity can cost thousands of dollars to identify and fix faults.



Due to this combinatorial explosion of potential root causes and high debugging costs, there is an urgent need for fault localization methods that, based on initial test case data, can identify a small number of suspicious input combinations (i.e., potential root causes) for further investigation. On this front, \cite{nie2011minimal} proposed a minimal failure-causing schema, which narrows down the search range for potential root causes and guides subsequent test case generation. \cite{niu2013identifying} proposed a notion of tuple relationship tree for visualizing the relationships among all input combinations. Such a tree is utilized to eliminate ``healthy'' combinations and to propose subsequent test cases for further system examination. \cite{gholamhosseinghandehari2016fault} and \cite{ghandehari2018combinatorial} introduced a two-phase approach for finding faulty statements in a software system.  \cite{lekivetz2018fault} proposed a deterministic ranking procedure that incorporates a domain-knowledge-guided weighting scheme.

Despite this literature, there remains a need for a principled \textit{statistical} framework for fault localization. Such a framework, implemented in a Bayesian fashion, provides three practical advantages for efficient and confident fault localization. First, it can accelerate the identification of a few key suspicious combinations from the combinatorially many potential root causes using test data, enabling efficient fault diagnosis with a small number of test runs. In particular, it permits the principled comparison of potential root causes with differing numbers of inputs, which addresses a key limitation of existing methods (see Section \ref{sec:mot}). Second, a Bayesian framework facilitates the integration of domain and/or structural knowledge from test engineers via its prior specification. Here, domain knowledge refers to prior information from historical data, expert judgment or system characteristics that raise or reduce the suspiciousness of certain input factors. Structural knowledge captures fault behavior from an understanding of the software's architecture and how components interact \citep{lekivetz2018fault,lekivetz2021testing}. A careful integration of such knowledge can accelerate the identification of root causes when faced with an overwhelming list of potential causes. Finally, a Bayesian framework enables principled risk quantification, by providing for each suspicious combination the posterior probability that it may be a root cause given test outcomes. Such probabilities support confident fault localization by guiding the required debugging effort. If only several combinations have high posterior probabilities, then debugging costs are expected to be low; if many combinations have non-negligible probabilities, then a massive debugging effort may be needed to fully investigate such combinations. Recent work \citep{landsberg2016probabilistic,landsberg2018doric} points to an increasing need for probabilistic metrics for risk-aware fault localization to gain deeper insights on software failure behavior.

We propose a new Bayesian Fault LOcalization (BayesFLo) framework that effectively achieves these three advantages. The main workhorse of BayesFLo is a new Bayesian model on root cause indicators over all possible input combinations, which facilitates easy integration of domain knowledge from test engineers. This model further embeds the desirable principles of combination hierarchy and heredity \citep{lekivetz2021testing}, which capture structural knowledge on software root causes. We show that the integration of such principles, derived from the well-known principles of effect hierarchy and heredity \citep{wu2011experiments} for analyzing factorial experiments, can accelerate the identification of root causes from limited test runs. A key challenge for Bayesian computation is the sheer number of considered combinations, which renders the computation of posterior root cause probabilities to be infeasible without careful manipulation. We develop a new algorithm for efficient computation of such posterior probabilities, leveraging recent tools from integer programming and graph representations. Finally, we demonstrate the practical advantages of BayesFLo over existing methods in two fault localization case studies, on the Traffic Alert and Collision Avoidance System (TCAS; \citealp{bradley1992simulation}) for aircraft collision avoidance, and the Vulnerable Road User (VRU; \citealp{Gladisch2020}) protection tests for safe autonomous driving.

The paper is organized as follows. Section \ref{sec:mot} outlines our motivating application on fault localization of TCAS,
as well as limitations of the state-of-the-art. Section \ref{sec:BayesianModel} presents the BayesFLo model, and describes the combination hierarchy and heredity principles embedded in its prior specification. Section \ref{sec:posterior} proposes a novel algorithm for efficient computation of posterior root cause probabilities for potential root cause combinations. Section \ref{sec:case_studies} explores the application of BayesFLo in two practical case studies on the fault localization of the TCAS system and the VRU protection tests.
Section \ref{sec:conclusion} concludes the paper.

\section{Motivating Application: Fault Localization of TCAS}
\label{sec:mot}

\subsection{Background \& Challenges}
\begin{figure}[!t]
    \centering
    \begin{subfigure}[b]{0.45\textwidth}
    \centering
        \includegraphics[width=0.99\linewidth]{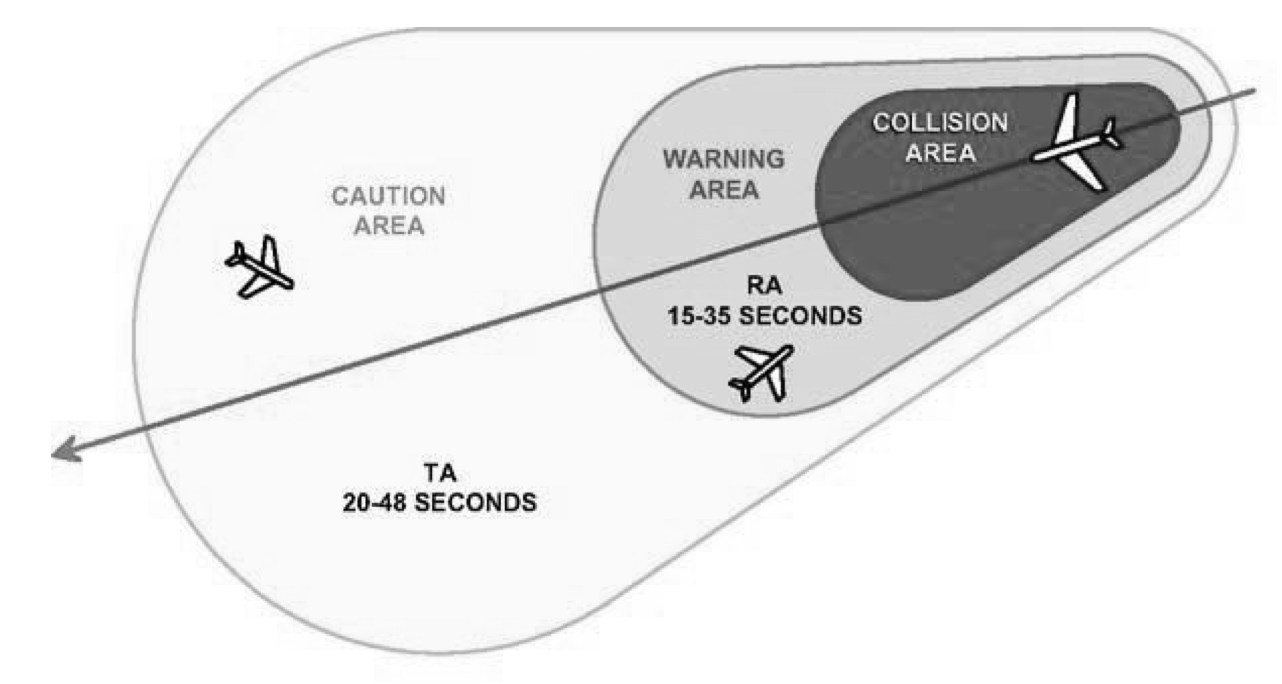}
    \end{subfigure}
    \begin{subfigure}[b]{0.45\textwidth}
    \centering
        \includegraphics[width=0.9\linewidth]{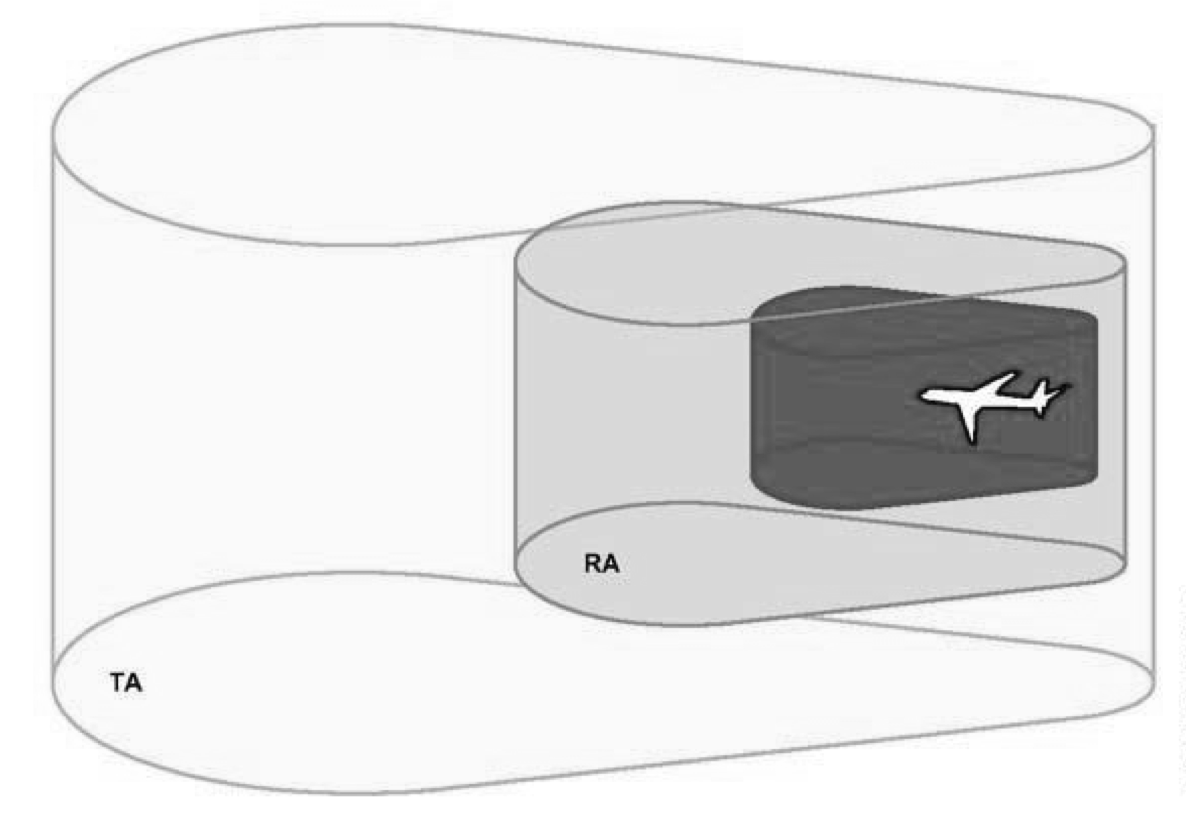}
    \end{subfigure}
    \caption{Visualizing the protection volume module in the TCAS software system \citep{TCASFAA}.}
    \label{fig:TCAS}
\end{figure}

Our motivating application is in the fault localization of the Traffic {Alert} and Collision Avoidance System (TCAS; \citealp{harman1989tcas,williamson1989development}), which is jointly developed by the Federal Aviation Administration (FAA) and the aviation industry. TCAS 
is designed to prevent potential mid-air collisions between aircraft \citep{TCASFAA}. Figure \ref{fig:TCAS} visualizes its operation. TCAS monitors the airspace around an aircraft, and determines a protection volume around the aircraft based on its speed and trajectory. When two aircraft come too close, TCAS alerts the pilots by issuing a resolution advisory. 
TCAS is currently mandated for U.S. commercial aircraft, and is also installed on many helicopters and turbine-powered general aviation aircraft \citep{TCASFAA}. Given the catastrophic implications of aircraft collisions, e.g., from recent events \citep{dccrash}, it is important to mitigate such risks by ensuring the reliability of TCAS systems 
via comprehensive testing under different scenarios.

\begin{figure}[!t]
    \centering
    \begin{tabular}{c c c }
        \toprule
        \textbf{Factor} & \textbf{Description} & \textbf{Levels} \\
        \hline
        1 & \texttt{Cur\_Vertical\_Sep} & 299, 300, 601\\
        2 & \texttt{High\_Confidence} & 0, 1\\
        3 & \texttt{Two\_of\_Three\_Reports\_Valid} & 0, 1\\
        4 & \texttt{Own\_Tracked\_Alt} & 1, 2\\
        5 & \texttt{Own\_Tracked\_Alt\_Rate} & 600, 601\\
        6 & \texttt{Other\_Tracked\_Alt} & 1, 2\\
        7 & \texttt{Alt\_Layer\_Value} & 0, 1, 2, 3\\
        8 & \texttt{Up\_Separation} & 0, 300, 399\\
        9 & \texttt{Down\_Separation} & 640, 740, 840\\
        10 & \texttt{Other\_RAC} & 0, 1, 2\\
        11 & \texttt{Other\_Capability} & 1, 2\\
        12 & \texttt{Climb\_Inhibit} & 0, 1\\
        \toprule
    \end{tabular}
    \captionof{table}{Input factors and their corresponding levels for our motivating TCAS case study.}
    \label{tab:TCAS_vars}
\end{figure}

Testing the TCAS software system 
has attracted much attention within the software engineering community, and there has been fruitful work on designing test suites for detecting defects in the system \citep{griesmayer2007automated,ghandehari2013applying}. In what follows, we consider a specific TCAS program from the TCAS software system
-- the Siemens test suite in \cite{griesmayer2007automated} -- for fault localization. This program takes in $I=12$ input factors (summarized in Table \ref{tab:TCAS_vars}), with each factor parameterizing certain characteristics of the aircraft system. Of these $I=12$ factors, nine pertain to the primary aircraft, and the remaining three (with prefix \texttt{Other} in Table \ref{tab:TCAS_vars}) are for an incoming aircraft that has entered its protection volume \citep{xie2013metamorphic}. The latter consists of inputs 6, 10 and 11, which reflect the altitude, resolution advisory and TCAS capability of the incoming aircraft. The former consists of remaining inputs that reflect the operating conditions of the primary aircraft, e.g., its current altitude, altitude rate-of-change, and separation altitudes with the incoming aircraft. 

 A fundamental challenge here is the expensive testing cost for each run of TCAS. Recall that a ``failure'' in a software program is when its actual behavior {deviates} from its expected behavior. Here, the assessment of the latter requires the construction of eleven data collection sites over the U.S. \citep{bradley1992simulation} and the collection of ground radar data (spanning hundreds of hours) for generating encounter databases. Logic test simulations are performed on thousands of such encounters \citep{bradley1992simulation}, and physical experiments are then conducted to validate its real-world performance. Such testing costs make the exhaustive testing of TCAS over all combinations of the $I=12$ input factors wholly impractical. 

An effective solution is to first construct a small designed set of inputs for testing. A reasonable design is a \textit{covering array} (CA; \citealp{colbourn2004combinatorial}), which covers each input combination (up to a certain order) at least once. Formally, take an array with dimensions $M \times I$, and suppose its $i$-th column takes $J_i$ distinct levels for some integer $J_i \geq 2$. Then this array is a CA of strength $s$, if for any choice of $s$ columns, each level combination involving such columns occurs at least once. For software testing, this CA can be used to design a test set with $M$ runs and $I$ inputs, where the $i$-th input has $J_i$ levels; the levels in its $m$-th row specify the input settings for the $m$-th test run. 
Table \ref{tab:TCAS_design} shows a strength-2 CA for our TCAS case study. 
This design achieves the desired coverage with
$M=19$ runs, which reduces the number of runs on the expensive software system. Using this CA, all single inputs and 2-factor input combinations are investigated in at least one test run; if there is a bug within such combinations, we would observe a corresponding failure in a test run.

The next step is to use these test outcomes to identify and rank potential root causes, i.e., fault localization. The last column in Table \ref{tab:TCAS_design} shows the outcomes (0 for a passed case, 1 for a failed case) for each test run. Here, a key challenge is that there are more than $\sum_{i=1}^{12}\binom{12}{i} 2^i > 500,000$ input combinations (e.g., the 2-factor combination 
\texttt{Cur\_Vertical\_Sep} = 299 and \texttt{High\_Confidence} = 0) that may be failure-inducing root causes. Since each combination is either a root cause or not, there are thus $2^{500,000}$ different root cause scenarios to consider for fault localization! The goal is to parse through these many scenarios to find which are likely given the observed test results. With the costly nature of debugging, this set of likely scenarios should further consist of \textit{only a few} combinations, to facilitate cost-efficient downstream diagnosis. Another challenge is the need to assess, for a suspicious combination, the probability this is indeed a root cause given test outcomes. This provides test engineers a principled way to gauge which combinations are likely root causes that need to be further investigated, and which can be confidently ignored. This probabilistic assessment of risk is desirable for trustworthy fault diagnosis \citep{landsberg2018doric,zhou2023trustworthy}. We inspect next existing methods through the lens of these two challenges.

\begin{figure}
    \centering
    \scalebox{0.95}{
    \begin{tabular}{c cccccccccccc c}
        \toprule
        & \multicolumn{12}{c}{\textbf{Factor}} &  \\
        \textbf{Run Number} & \textbf{1} & \textbf{2} & \textbf{3} & \textbf{4} & \textbf{5} & \textbf{6} & \textbf{7} & \textbf{8} & \textbf{9} & \textbf{10} & \textbf{11} & \textbf{12} & \textbf{Outcome}\\
        \hline
        1&601&1&1&1&601&1&3&300&640&1&1&0  &0\\
        2&300&1&1&1&600&1&0&0&740&0&1&1  &0\\
        3&299&0&0&2&600&1&2&300&640&0&1&0  &0\\
        4&300&0&0&2&600&2&1&300&740&2&1&0  &0\\
        5&601&0&0&2&601&1&3&399&840&1&1&1  &0\\
        6&601&1&0&1&601&2&0&300&840&0&2&0  &0\\
        7&300&1&0&2&600&2&3&0&840&0&2&0  &0\\
        8&299&0&1&2&600&1&1&0&640&2&1&1  &0\\
        9&299&1&0&1&601&2&0&399&640&2&1&0  &0\\
        10&601&1&1&1&601&1&1&0&740&2&2&1  &0\\
        11&299&0&0&2&600&1&3&399&740&0&2&1  &0\\
        12&300&1&1&2&601&1&2&399&840&2&2&1  &0\\
        13&299&0&1&1&601&1&1&399&640&0&1&1  &1\\
        14&299&0&1&1&600&1&1&300&840&1&1&1  &0\\
        15&300&0&0&2&600&1&0&399&640&1&2&1  &1\\
        16&601&1&1&1&600&2&2&0&740&1&1&1  &0\\
        17&299&0&0&1&601&1&2&399&640&2&1&0  &0\\
        18&299&0&1&1&600&1&1&300&840&0&1&1  &0\\
        19&601&1&0&1&600&1&3&0&640&2&1&0  &0\\
        \toprule
    \end{tabular}}
    \captionof{table}{The $M=19$-run CA design and test outcomes for our motivating TCAS case study. Here, an outcome of 0 indicates a passed test case, with 1 indicating a failed test case.}
    \label{tab:TCAS_design}
\end{figure}

\subsection{State-of-the-Art and Its Limitations}



Section \ref{sec:Intro} provides a literature review of existing fault localization methods, which we do not repeat here for brevity. We review next two state-of-the-art techniques, the BEN approach in \cite{ghandehari2018combinatorial} and the weighting approach in \cite{lekivetz2018fault}, which will serve as our benchmarks later.

The BEN approach in \cite{ghandehari2018combinatorial} is a two-phase method, where Phase 1 identifies and ranks potential root causes and Phase 2 locates faulty statements. Phase 1 is most relevant for our fault localization goal, as Phase 2 considers the downstream problem of debugging the software system. Phase 1 of BEN employs the so-called component suspiciousness metric, defined for a component $(i,j)$ (i.e., level $j$ of factor $i$) as $\rho(i,j) = \frac{1}{3} [ u(i,j) + v(i,j) + w(i,j)]$. Here, $u(i,j)$ denotes the ratio of failed test cases with component $(i,j)$ over the total number of failed test cases, $v(i,j)$ denotes the ratio of failed test cases with $(i,j)$ over the total number of test cases with $(i,j)$, and $w(i,j)$ denotes the ratio of suspicious combinations with $(i,j)$ over the total number of suspicious combinations. The suspiciousness metric of an input combination is defined as the average component suspiciousness of its components. This metric is then used (along with a measure of environmental suspiciousness) to rank the suspiciousness of potential root causes from test outcomes.

One drawback with BEN is that it cannot incorporate domain knowledge from test engineers. To address this, \cite{lekivetz2018fault} adopts an alternate weighting approach. Here, a weight is first assigned to each factor level: the baseline weight is 1, and a higher weight is assigned if one believes (e.g., from expert judgment or system characteristics) that such a level is more likely to cause a failure. The weight of an input combination is taken as the product of the weights for its components. Suspicious combinations are then ranked based on their ``normalized'' weights. In the simple case where each failed run involves a separate set of potential root causes, the normalized weight of a suspicious combination is computed as its weight divided by the sum of weights for all potential root causes for its corresponding failed run. When there are overlaps in potential root causes between different failed runs, a complete enumeration of potential root cause scenarios is needed to compute the normalized weights. This enumeration is possible via brute force with a small number of failed test runs. However, as the number of failed runs grows, such an enumeration quickly becomes computationally infeasible; more on this in Section \ref{sec:vru}.

Note that, for this weighting approach, the weight of a higher-order combination (e.g., factors A, B, C at level 1) is by construction larger than that for a lower-order combination (e.g., factors A, B at level 1). Thus, applying this weighting approach for combinations of different orders may lead to unintended heightened suspiciousness of higher-order combinations, which is undesired. To avoid this, \cite{lekivetz2018fault} recommends that this approach be used to iteratively analyze combinations of the same order, beginning with lower-order combinations (which have greater suspiciousness a priori by the principle of combination hierarchy; see \citealp{lekivetz2021testing}). In other words, combinations of order 2 should be analyzed and debugged first, then combinations of order 3, and so on. The same iterative analysis procedure over combination order is employed for the BEN approach in \cite{ghandehari2018combinatorial} for a similar reason.


We now investigate these methods for the motivating TCAS application. Here, the true root cause -- a fully-documented fault in \cite{griesmayer2007automated} -- is a single 3-factor combination involving factors 12, 8 and 9; this fault is further discussed in Section \ref{sec:TCAS}. The BEN approach in \cite{ghandehari2018combinatorial} is implemented following the algorithm in their paper. For the weighting approach in \cite{lekivetz2018fault}, we incorporate domain knowledge on the nature of the underlying fault via higher weights (i.e., heightened suspiciousness) on certain factors; details are provided later in Section \ref{sec:TCAS}. Since there are only two failed test runs here, its normalized weights can be computed via a brute-force enumeration of potential root cause scenarios. Table \ref{tab:TCAS_RefRes} shows the ranked list of 2-factor and 3-factor combinations (2-FCs and 3-FCs) identified by these two approaches, with the true 3-factor root cause bolded.


This table reveals two notable limitations of these existing methods. First, since both methods employ an iterative analysis approach with respect to combination order, they require the investigation of all eight 2-FCs before finding the true 3-factor root cause. Given the expensive nature of debugging each combination (which can take thousands of dollars), this considerably increases the cost of identifying the true root cause. A key goal is thus to identify the true root cause with as few such follow-up investigations as possible \citep{landsberg2018doric}. A Bayesian approach can achieve this via a probabilistic modeling framework over all combination orders, which permits a fair comparison of suspicious combinations across orders via its posterior probabilities. Second, while both approaches employ different suspiciousness metrics for ranking combinations, such metrics are not posterior probabilities and do not provide a principled quantification of risk for each combination. This can hinder actionable decisions: here, it is not clear from existing analyses whether one should, e.g., investigate all 3-FCs, or whether one can confidently ignore the lower-ranked combinations. A Bayesian approach can address this limitation, by providing posterior probabilities on suspicious combinations (along with a user-specified significance level) that can help plan the debugging effort needed for \textit{confident} fault localization.

\begin{figure}[!t]
  \centering
  \begin{subfigure}[t]{0.48\textwidth}
    \centering
    \scalebox{0.8}{
    \begin{tabular}{ccc}
      \toprule
      \textbf{Combination Type} & \textbf{Count} & \textbf{Normalized Weight}\\
      \midrule
      2-FC & 1 & 1.00 \\
      2-FC & 2 & 0.22\\
      2-FC & 5 & 0.11\\
      \midrule
      \textbf{3-FC} & \textbf{1} & \textbf{0.65}\\
      3-FC & 140 & $<$ 0.02\\
      \bottomrule
    \end{tabular}}
    \caption{Weighting approach.}
  \end{subfigure}\hfill
  \begin{subfigure}[t]{0.48\textwidth}
    \centering
    \scalebox{0.8}{
    \begin{tabular}{ccc}
      \toprule
      \textbf{Combination Type} & \textbf{Count} & \textbf{Rank}\\
      \midrule
      2-FC & 2 & 1\\
      2-FC & 6 & 3-8\\
      \midrule
      \textbf{3-FC} & \textbf{1} & \textbf{1}\\
      3-FC & 140 & 2-141\\
      \bottomrule
    \end{tabular}}
    \caption{BEN (Phase 1).}
  \end{subfigure}
  \captionof{table}{Suspicious 2-FCs and 3-FCs ranked using the weighting approach in \cite{lekivetz2018fault} and the BEN approach (Phase 1) in \cite{ghandehari2018combinatorial} for our motivating TCAS case study. Combinations at the top are deemed more suspicious and are investigated first. The true root cause is \textbf{bolded}.}
  \label{tab:TCAS_RefRes}
\end{figure}

\section{The BayesFLo Model}
\label{sec:BayesianModel}

We propose a new Bayesian Fault Localization (BayesFLo) framework, which provides a principled probabilistic approach for fault localization via conditioning on test set outcomes. We first present this Bayesian model, then show how it permits the integration of domain and structural knowledge within its prior specification.


Here, it is worth mentioning that, while there may be different ``types'' of failures (e.g., incorrect or missing outputs), the system has typically undergone a reasonable level of preliminary testing (e.g., unit testing, which eliminates root causes in single units) at time of combinatorial testing. Because of this, there remains only a very limited number of faults for diagnosis \citep{lekivetz2021testing}. As such, we presume in the following only a single failure type at time of BayesFLo analysis. We will briefly discuss an approach for the rare scenario of multiple failure types later in the Conclusion.


\subsection{Prior Specification}
\label{sec:prior}



We first introduce some notation. Consider a software system (or more broadly, a complex engineering system) with $I \geq 1$ input factors, where a factor $i$ can take on $J_i \geq 2$ different levels. 
A $K$-factor combination (with $K\leq I$) is denoted as $(\mathbf{i},\mathbf{j})_K$, where $\mathbf{i} = (i_1, \cdots, i_K), i_1 < \cdots < i_K$ is an ordered $K$-vector containing all inputs for this combination, and $\mathbf{j} = (j_1,\cdots,j_K), j_k \in \{1, \cdots, J_{i_k}\}$ is a $K$-vector indicating the levels of each corresponding input. For example, the 2-factor combination of the first factor at level 1 and the second factor at level 2 can be denoted as $(\mathbf{i},\mathbf{j})_2$, where $\mathbf{i} = (1,2)$ and $\mathbf{j} = (1,2)$. In the case of $K=1$, i.e., a single input $i$ at level $j$, this may be simplified to $(i,j)$. 



Now let $\mathcal{C}_K$ denote the set of $K$-factor combinations $(\mathbf{i},\mathbf{j})_K$ as described above, and let $\mathcal{C} = \cup_{K=1}^I \mathcal{C}_K$ denote the set of combinations over all orders $K = 1, \cdots, I$. Further let $Z_{(\mathbf{i},\mathbf{j})_K} \in \{0,1\}$ be an indicator variable for whether the combination $(\mathbf{i},\mathbf{j})_K$ is truly a root cause. As this is unknown prior to running test cases, we model each $Z_{(\mathbf{i},\mathbf{j})_K}$ a priori as an independent Bernoulli random variable of the form:
\begin{equation}
Z_{(\mathbf{i},\mathbf{j})_K}\overset{indep.}{\sim} \text{Bern}\{p_{(\mathbf{i},\mathbf{j})_K}\}, \quad (\mathbf{i},\mathbf{j})_K \in \mathcal{C}_K, \quad K = 1, \cdots, I,
\label{eq:bern}
\end{equation}
where $p_{(\mathbf{i},\mathbf{j})_K}$ is the prior probability that this combination is a root cause. Here, the view that $Z_{(\mathbf{i},\mathbf{j})_K}$ is random makes our approach Bayesian; this contrasts with existing fault localization approaches, which presume $Z_{(\mathbf{i},\mathbf{j})_K}$ to be fixed but unknown. For $K=1$, this notation simplifies to $Z_{(i,j)}$ and $p_{(i,j)}$. Whenever appropriate, we denote $\mathbf{Z} = (Z_{(\mathbf{i},\mathbf{j})_K})_{(\mathbf{i},\mathbf{j})_K \in \mathcal{C}}$ and $\mathbf{p} = (p_{(\mathbf{i},\mathbf{j})_K})_{(\mathbf{i},\mathbf{j})_K \in \mathcal{C}}$ for brevity.

It is worth noting the sheer number of input combinations in $\mathcal{C}$ that needs to be considered as potential root causes. Assuming each factor has an equal number of levels $J = J_1 = \cdots = J_I$, one can show that $\mathcal{C}_K$ contains $\binom{I}{K}J^K$ distinct combinations of order $K$, thus the total number of considered input combinations is $|\mathcal{C}| = $ $\sum_{K=1}^I\binom{I}{K}J^K$. Even with a moderate number of inputs, say $I=10$, with each having $J=2$ levels, this amounts to $|\mathcal{C}| = 59,048$ combinations. As we shall see later in Section \ref{sec:posterior}, the size of $\mathcal{C}$ forms the key bottleneck for Bayesian inference, as the computation of posterior probabilities can require $\mathcal{O}(2^{|C|})$ work; this can thus be infeasible even for small software systems.




Next, we adopt the following product form on the root cause probability for $(\mathbf{i},\mathbf{j})_K$:
\begin{eqnarray}
    p_{(\mathbf{i},\mathbf{j})_K} = \prod_{k=1}^K  p_{(i_k,j_k)}, \quad (\mathbf{i},\mathbf{j})_K \in \mathcal{C}_K, \quad K = 2, \cdots, I.
    \label{eqn:prior_propagation}
\end{eqnarray}
In words, the combination root cause probability $p_{(\mathbf{i},\mathbf{j})_K}$ is modeled as the {product} of the root cause probabilities for its component inputs. The key advantage of this form is that it captures \textit{structural knowledge} on root causes, by embedding the principles of combination hierarchy and heredity \citep{lekivetz2021testing}. These principles reflect the structured nature of typical software root causes, and can be seen as extensions of the well-known principles of effect hierarchy and heredity \citep{wu2011experiments}, which are widely used for analysis of factorial experiments. The first principle, combination hierarchy, asserts that combinations involving fewer inputs are more likely to be failure-inducing than those involving more inputs. Empirical evidence suggests this principle holds for software systems across various domains \citep{kuhn2004software}. To see how our prior \eqref{eqn:prior_propagation} captures combination hierarchy, note that by its product form construction, the combination probability $p_{(\mathbf{i},\mathbf{j})_K}$ is always less than the probability of any component input $p_{(i_k,j_k)}$. Thus, this prior assigns increasingly smaller root cause probabilities on combinations with a higher combination order $K$, thus capturing the desired hierarchy structure. The second principle, combination heredity, asserts that a combination is more likely to be failure-inducing when some of its component inputs are more likely to be failure-inducing. From our product-form prior \eqref{eqn:prior_propagation}, note that the combination root cause probability $p_{(\mathbf{i},\mathbf{j})_K}$ cannot be large unless some of its component root cause probabilities in $\{p_{(i_k,j_k)}\}_{k=1}^K$ are also large. This thus captures the desired combination heredity effect. Similar product-form weights have been used for capturing hierarchy and heredity for predictive modeling \citep{tang2023hierarchical} and data reduction \citep{mak2017projected}.

With the product-form prior \eqref{eqn:prior_propagation}, we require only the specification of the single-input root cause probabilities $\{p_{(i,j)}\}_{i,j}$, which can be used to integrate further \textit{domain knowledge} from test engineers. For most software systems at the testing stage, it may be reasonable to specify a small (i.e., near-zero) value for $p_{(i,j)}$, as this reflects the prior belief that failure-inducing root causes should occur sporadically. Oftentimes, however, an engineer has additional domain knowledge that permits a more informed prior specification, e.g., from historical data or expert judgment. For example, the engineer may know that certain factors have been recently added to the system, and thus may be more suspicious of such factors. This heightened suspicion can be captured via a larger specification of its $p_{(i,j)}$ compared to other factors. We shall see how such domain knowledge can improve fault localization in later case studies.


From a Bayesian perspective, the product-form prior \eqref{eqn:prior_propagation} provides a way for propagating the elicited domain knowledge over the many root cause probabilities in $\mathbf{p}$. For example, suppose an engineer has heightened suspicions on factor $i$, and accordingly specifies a higher value for the single-factor prior probabilities $\{p_{(i,j)}\}_j$. By \eqref{eqn:prior_propagation}, this induces larger prior root cause probabilities $p_{\ijbold_K}$ for any combination $\ijbold_K$ involving factor $i$, thus ``pooling'' this information over such combinations. This ``information pooling'', guided by the embedded principles of combination hierarchy and heredity, can facilitate the disentangling of the large number of potential root causes from limited test runs. Recent work on related notions of information pooling have shown promise in high-dimensional inference problems, e.g., matrix completion \citep{yuchi2023bayesian} and multi-armed bandits \citep{mak2022tsec}; we show that this is also similarly important for efficient fault localization. Of course, as with any Bayesian modeling framework, the domain and/or structural knowledge captured within the BayesFLo prior is not a strict assumption; it can be overturned with sufficient evidence from data.

\subsection{Posterior Root Cause Probabilities}
\label{sec:likelihood}
In what follows, we suppress the notation $\ijbold_K$ to $\ijbold$ for brevity. Using the above prior specification, we now need to condition on the test outcome data. Suppose we run the software system at $M$ different test cases, where the $m$-th test case is performed at input levels $\mathbf{t}_m = (t_{m,1}, \cdots, t_{m,I})$, $t_{m,i} \in \{1, \cdots, J_i\}$. Then the test data can be denoted as $\mathcal{D} = \{(\mathbf{t}_m,y_m)\}_{m=1}^M$, where $y_m \in \{0,1\}$ is a binary variable with 1 indicating a failure and 0 if not. To make things concrete, consider the following example. Suppose the system has $I=3$ input factors, each with two levels. Further assume there is only one true root cause $((1,2),(1,2))$, i.e., the combination of the first input at level 1 and the second at level 2, which results in failure. Suppose we then run the first test case at input setting $\mathbf{t}_1 = (1,2,1)$, i.e., with the three factors at levels 1, 2 and 1, respectively. Then, since the root cause is present in $\mathbf{t}_1$, this would result in a failure, namely $y_1 = 1$. However, if we run the second test case at a different setting $\mathbf{t}_2 = (2,2,2)$, then this test case would result in no failure, i.e., $y_2 = 0$, as the root cause is not present in $\mathbf{t}_2$. Here, we presume that observed outcomes are \textit{deterministic}, in that the same outcome $y_m$ is always observed whenever the software system is run with inputs $\mathbf{t}_m$.

With this framework, the problem of fault localization then reduces to the evaluation of the posterior root cause probabilities for \textit{all} considered combinations in $\mathcal{C}$, namely:
\begin{equation}
\small
\mathbb{P}(Z_{(\mathbf{i},\mathbf{j})}=1 | \mathcal{D}), \quad  \text{for all} \;(\mathbf{i},\mathbf{j}) \in \mathcal{C}.
\label{eq:postprob}
\end{equation}
\normalsize
Such a computation, however, can easily become computationally intractable. The key bottleneck lies in the complex conditioning structure from data $\mathcal{D}$ over the {high-dimensional} set of combinations $\mathcal{C}$; as we see later, this can then induce an $\mathcal{O}(2^{|\mathcal{C}|})$ complexity for a brute-force computation of posterior probabilities. Recall that, with the moderate setting of $I=10$ and $J=2$, $|\mathcal{C}|$ consists of nearly 60,000 combinations. Thus, without careful modifications to exploit problem structure, posterior computation can be intractable even for small systems.

\begin{figure}
    \centering
    \includegraphics[width=0.5\linewidth]{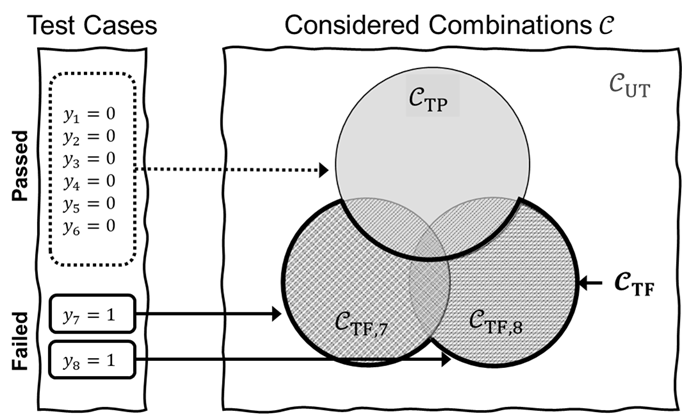}
\caption{Visualizing the use of passed and failed test cases for partitioning the set of considered combinations $\mathcal{C}$ into $\mathcal{C}_{\rm TP}$, $\mathcal{C}_{\rm TF}$ and $\mathcal{C}_{\rm UT}$.\vspace{0.3cm}}

    \label{fig:Combination_partition_workflow}
\end{figure}

We adopt next the following categorization of input combinations in $\mathcal{C}$ for efficient computation of root cause probabilities:
\begin{enumerate}[(a)]
    \item \textbf{Tested-and-Passed (TP)}: TP combinations for a \textit{passed} test case $\mathbf{t}_m$ are combinations in $\mathcal{C}$ that have been {tested} in $\mathbf{t}_m$. Continuing from the earlier example, suppose we run the test case $\mathbf{t}_m = (2,2,2)$ with no failure, i.e., with $y_m = 0$. Then it follows that the combination $((1,2),(2,2))$, i.e., with the first factor A at level 2 and the second factor B at level 2, is a TP combination. (In what follows, we may denote such a combination as $\text{A}_2\text{B}_2$ for notational simplicity; this should be clear from context.) For this single passed case, the set of TP combinations is $\mathcal{C}_{{\rm TP},m} =\{\text{A}_2,\text{B}_2,\text{C}_2,\text{A}_2\text{B}_2,\text{A}_2\text{C}_2,\text{B}_2\text{C}_2,\text{A}_2\text{B}_2\text{C}_2\}$. 
    \item \textbf{Tested-and-Failed (TF)}: TF combinations for a \textit{failed} test case $\mathbf{t}_m$ are combinations in $\mathcal{C}$ that have been {tested} in $\mathbf{t}_m$. For example, suppose we run the test case $\mathbf{t} = (1,2,1)$ and observe a failure, i.e., with $y_m = 1$. Then, from this single failed case, the set of TF combinations becomes $\mathcal{C}_{{\rm TF},m} = \{\text{A}_1,\text{B}_2,\text{C}_1,\text{A}_1\text{B}_2,\text{A}_1\text{C}_1,\text{B}_2\text{C}_1,\text{A}_1\text{B}_2\text{C}_1\}$.
    \item \textbf{Untested (UT)}: UT combinations are combinations in $\mathcal{C}$ that have \textit{not} been tested in any test case. For example, suppose we run the test case $\mathbf{t} = (1,2,1)$. Then one UT combination is $\text{A}_2\text{B}_1$, as such a combination was not tested in $\mathbf{t}$.
\end{enumerate}
This partition of $\mathcal{C}$ naturally extends for multiple test runs in $\mathcal{D}$. Here, the TP combinations $\mathcal{C}_{\rm TP}$ from $\mathcal{D}$ are the TP combinations over all \textit{passed} test cases. The TF combinations $\mathcal{C}_{\rm TF}$ from $\mathcal{D}$ are the TF combinations over all \textit{failed} test cases, with the combinations from $\mathcal{C}_{\rm TP}$ removed. $\mathcal{C}_{\rm UT}$ then consists of all remaining combinations in $\mathcal{C}$. In other words:
\begin{equation}
\label{eq:classify}
\mathcal{C}_{\rm TP} = \cup_{m:y_m = 0} \; \mathcal{C}_{{\rm TP},m}, \quad \mathcal{C}_{\rm TF} = (\cup_{m:y_m = 1} \; \mathcal{C}_{{\rm TF},m}) \setminus \mathcal{C}_{\rm TP}, \quad \mathcal{C}_{\rm UT} = \mathcal{C} \setminus (\mathcal{C}_{\rm TP} \cup \mathcal{C}_{\rm TF}).
\end{equation}
Figure \ref{fig:Combination_partition_workflow} visualizes this partition of $\mathcal{C}$ from observed test runs for a simple example.

With this partition of $\mathcal{C}$, we now investigate efficient algorithms for computing the posterior root cause probabilities \eqref{eq:postprob}. For TP combinations, it is clear that such a combination cannot be a root cause, as it was cleared by a passed test case. In other words:
\begin{equation}
\mathbb{P}(Z_{(\mathbf{i},\mathbf{j})}=1 | \mathcal{D}) = 0, \; (\mathbf{i},\mathbf{j}) \in \mathcal{C}_{\rm TP}.
\label{eq:postprobtp}
\end{equation}
This is akin to \cite{ghandehari2018combinatorial}, which removes TP combinations from consideration for root causes. Furthermore, for UT combinations, we have:
\begin{equation}
\mathbb{P}(Z_{(\mathbf{i},\mathbf{j})}=1 | \mathcal{D}) = \mathbb{P}(Z_{(\mathbf{i},\mathbf{j})}=1 ), \quad (\mathbf{i},\mathbf{j}) \in \mathcal{C}_{\rm UT},
\label{eq:postprobut}
\end{equation}
since the observed test set $\mathcal{D}$ does not provide any information on an untested combination $\ijbold$. As such, its root cause probability given $\mathcal{D}$ simply reduces to its prior probability given in \eqref{eqn:prior_propagation}. The challenge thus lies in computing posterior probabilities on the remaining class of TF combinations. We present next an approach for computing such probabilities, leveraging tools from integer programming and graph representations.

\section{Computation of Posterior Root Cause Probabilities}
\label{sec:posterior}


Consider the case of TF combinations, where we wish to compute the posterior root cause probability \eqref{eq:postprob} for a given TF combination $(\mathbf{i},\mathbf{j}) \in \mathcal{C}_{\rm TF}$. One solution might be the following ``brute-force'' approach:
\begin{equation}
\mathbb{P}(Z_{(\mathbf{i},\mathbf{j})}=1 | \mathcal{D}) = \frac{\mathbb{P}(Z_{(\mathbf{i},\mathbf{j})}=1,\mathcal{D})}{\mathbb{P}(\mathcal{D})} = \frac{\sum_{\mathbf{z} 
\in \{0,1\}^{|\mathcal{C}|}, Z_{(\mathbf{i},\mathbf{j})}=1}\mathbb{P}(\mathbf{Z} = \mathbf{z})\mathbb{P}(\mathcal{D}|\mathbf{Z} = \mathbf{z})}{\sum_{\mathbf{z} 
\in \{0,1\}^{|\mathcal{C}|}}\mathbb{P}(\mathbf{Z} = \mathbf{z})\mathbb{P}(\mathcal{D}|\mathbf{Z} = \mathbf{z})}.
\label{eq:brute}
\end{equation}
where $\mathbb{P}(\mathbf{Z} = \mathbf{z})$ follows from Equation \eqref{eqn:prior_propagation}, and $\mathbb{P}(\mathcal{D}|\mathbf{Z} = \mathbf{z})$ can be deduced from the reasoning in Section \ref{sec:likelihood}. The limitation of such an approach is clear. For each $(\mathbf{i},\mathbf{j}) \in \mathcal{C}_{\rm TF}$, we need to compute the sum of $2^{|\mathcal{C}|-1}$ terms in the numerator and the sum of $2^{|\mathcal{C}|}$ terms in the denominator. Hence, even for small systems with $|\mathcal{C}|$ small, this brute-force approach can be infeasible. This sheer dimensionality of potential root cause scenarios is a key bottleneck for tractable computation of posterior probabilities for Bayesian fault localization.


To address this, we employ an alternate reformulation, which permits considerable speed-ups for computing probabilities. We first outline this reformulation, then show how this facilitates efficient computation via a connection to the related problem of minimal set covering.

\subsection{An Alternate Formulation}
The following proposition provides a useful reformulation of the desired posterior root cause probability for a TF combination $(\mathbf{i},\mathbf{j})$:
\vspace{-0.2cm}
\begin{prop}
Let $(\mathbf{i},\mathbf{j}) \in \mathcal{C}_{\rm TF}$, and let:
\begin{equation}
\mathcal{M}_{(\mathbf{i},\mathbf{j})} = \{m = 1, \cdots, M : y_m = 1, (\mathbf{i},\mathbf{j}) \in \mathcal{C}_{{\rm TF},m}\}
\end{equation}
be the index set of failed test cases for which $(\mathbf{i},\mathbf{j})$ is a potential root cause. Define the event: 
\begin{equation}
E_{(\mathbf{i},\mathbf{j})}=\{\textup{for each $m \in \mathcal{M}_{(\mathbf{i},\mathbf{j})}$, there exists some $c \in \mathcal{C}_{{\rm TF},m} \setminus \mathcal{C}_{{\rm TP}}  $ such that $Z_c = 1$} \}.
\label{eq:eij}
\end{equation}
In words, this is the event that \textup{all} failures in $\mathcal{M}_{(\mathbf{i},\mathbf{j})}$ can be explained by the selected root causes $\{c \in \mathcal{C}_{\rm TF}: Z_c = 1\}$. The desired posterior root cause probability then reduces to:
\begin{equation}
\mathbb{P}(Z_{(\mathbf{i},\mathbf{j})}=1 | \mathcal{D}) = \mathbb{P}(Z_{(\mathbf{i},\mathbf{j})}=1 | E_{(\mathbf{i},\mathbf{j})}) = \frac{p_{(\mathbf{i},\mathbf{j})}}{\mathbb{P}(E_{(\mathbf{i},\mathbf{j})})}. 
\label{eq:altform}
\end{equation}
\end{prop}
\noindent The proof of this proposition can be found in Appendix A. There are two key advantages of the alternate form \eqref{eq:altform} over the brute-force approach \eqref{eq:brute}. First, its numerator can be directly computed via Equation \eqref{eqn:prior_propagation} with little work. Second, its denominator $\mathbb{P}(E_{(\mathbf{i},\mathbf{j})})$ can be efficiently computed via a novel connection to a related minimal set covering problem for bipartite graphs \citep{asratian1998bipartite}, which we show below.

\begin{figure}
    \centering
    \includegraphics[width=0.55\linewidth]{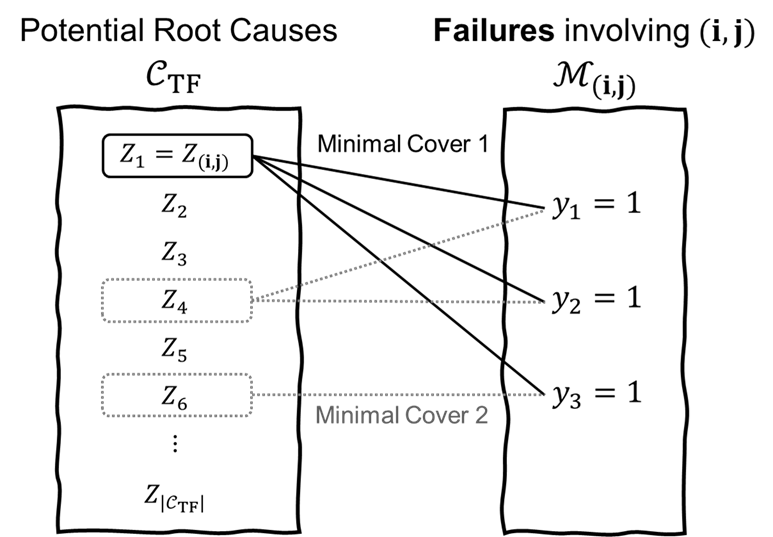}
    \caption{Visualizing the bipartite graph representation and two minimal covers for failures involving the combination $(\mathbf{i},\mathbf{j})$.}
    \label{fig:bipartite_graph_DAG}
\end{figure}

To compute $\mathbb{P}(E_{(\mathbf{i},\mathbf{j})})$, we first inspect condition \eqref{eq:eij} for $E_{(\mathbf{i},\mathbf{j})}$, which requires, for each failed test case in $\mathcal{M}_{(\mathbf{i},\mathbf{j})}$, a corresponding TF combination that induces this failure. Figure \ref{fig:bipartite_graph_DAG} visualizes this condition in the form of a bipartite graph, where the left nodes are the TF combinations in $\mathcal{C}_{\rm TF}$, and the right nodes are failed test cases in $\mathcal{M}_{(\mathbf{i},\mathbf{j})}$. Here, an edge is drawn from a combination $c$ (on left) to a test case index $m$ (on right) if $c \in \mathcal{C}_{{\rm TF},m}$, i.e., if combination $c$ is contained in the failed test inputs $\mathbf{t}_m$. Viewed this way, condition \eqref{eq:eij} is equivalent to finding a selection of potential root causes in $\{Z_c\}_{c \in \mathcal{C}_{\rm TF}}$, such that every failed test case on the right is connected to (or ``covered'' by) a selected combination on the left via an edge. Figure \ref{fig:bipartite_graph_DAG} visualizes two possible ``covers''. Such a cover of right-hand nodes can be interpreted as a selection of potential root causes (left-hand nodes) that \textit{explains} the failed test cases. Thus, to compute the probability $\mathbb{P}(E_{(\mathbf{i},\mathbf{j})})$, we need to sum over the prior probabilities for all possible selections of potential root causes that cover the failed test cases in $\mathcal{M}_{(\mathbf{i},\mathbf{j})}$.

\subsection{Enumerating Minimal Covers}
Using this insight, we now establish a useful link between the desired probability $\mathbb{P}(E_{(\mathbf{i},\mathbf{j})})$ and the related problem of minimal set covering. Formally, we define a \textit{cover} of the failed test indices $\mathcal{M}_{(\mathbf{i},\mathbf{j})}$ as a subset $\tilde{\mathcal{C}}$ of the potential root causes $\mathcal{C}_{\rm TF}$, such that for every $m \in \mathcal{M}_{(\mathbf{i},\mathbf{j})}$, there exists an edge connecting some node in $\tilde{\mathcal{C}}$ to $m$. A \textit{minimal cover} of $\mathcal{M}_{(\mathbf{i},\mathbf{j})}$ is then a cover $\tilde{\mathcal{C}}$ of $\mathcal{M}_{(\mathbf{i},\mathbf{j})}$ which, if any element is removed from $\tilde{\mathcal{C}}$, ceases to be a cover. Figure \ref{fig:bipartite_graph_DAG} visualizes this notion of a minimal cover.


With this definition, the following proposition reveals a useful connection:
\begin{prop}
The desired probability $\mathbb{P}(E_{(\mathbf{i},\mathbf{j})})$ can be simplified as:
\begin{equation}
\mathbb{P}(E_{(\mathbf{i},\mathbf{j})}) = \mathbb{P}(\{Z_c = 1 \textup{ for all } c \in \tilde{\mathcal{C}}\} \textup{, for at least one minimal cover $\tilde{\mathcal{C}}$ of $\mathcal{M}_{(\mathbf{i},\mathbf{j})}$}).
\label{eq:cover}
\end{equation}
\label{prop:cover}
\end{prop}
\vspace{-1cm}
\noindent Its proof can be found in Appendix B. This shows that $\mathbb{P}(E_{(\mathbf{i},\mathbf{j})})$ is equivalent to the probability that, for at least one minimal cover $\tilde{\mathcal{C}}$, all combinations in $\tilde{\mathcal{C}}$ are indeed root causes. 

To compute \eqref{eq:cover}, a natural approach is to first enumerate all minimal covers of $\mathcal{M}_{(\mathbf{i},\mathbf{j})}$. Fortunately, the set cover problem for bipartite graphs has been well-studied in the literature, and efficient polynomial-time algorithms have been developed for finding minimal covers \citep{skiena1998algorithm,hopcroft1973n}. Leveraging such developments can thus greatly speed up the brute-force approach for posterior probability computation (see Equation \eqref{eq:brute}), which is doubly-exponential in complexity and thus infeasible for even small software systems. With recent developments in integer programming algorithms \citep{wolsey2020integer}, a popular strategy for finding minimal set covers is to formulate and solve this problem as an integer linear program (ILP; \citealp{schrijver1998theory}). We adopt such a strategy below.


Let $\mathcal{C}_{\text{TF},(\mathbf{i},\mathbf{j})}$ be the set of potential root causes in $\mathcal{C}_{\rm TF}$ involving $\mathcal{M}_{(\mathbf{i},\mathbf{j})}$; this is typically much smaller than $\mathcal{C}_{\rm TF}$, which reduces the size of the optimization program below. We propose the following feasibility program to find one minimal cover for $\mathcal{M}_{(\mathbf{i},\mathbf{j})}$:
\begin{equation}
\label{eqn:ILP}
\resizebox{0.92\linewidth}{!}{%
$
\begin{split}
\argmax \; 1 \text{\quad s.t. \quad } &z_c \in \{0, 1\} \; \text{for all } c \in \mathcal{C}_{\text{TF},(\mathbf{i},\mathbf{j})}, \quad l_{g,m} \in \{0,1\} \; \text{for all } g \in \mathcal{C}_{\text{TF},(\mathbf{i},\mathbf{j})}, m\in\mathcal{M}_{(\mathbf{i},\mathbf{j})},\\
    \text{[C1] }& \sum_{c \in \mathcal{C}_{\text{TF},(\mathbf{i},\mathbf{j})}}z_c\cdot \mathbb{I}(m\in\mathcal{M}_c)\geq 1\; \text{for all } m\in\mathcal{M}_{(\mathbf{i},\mathbf{j})},\\
    \text{[C2] }& \sum_{c \in \mathcal{C}_{\text{TF},(\mathbf{i},\mathbf{j})},c\neq g}z_c\cdot\mathbb{I}(m\in\mathcal{M}_c)\leq |\mathcal{C}_{\text{TF},(\mathbf{i},\mathbf{j})}|(1-l_{g,m}) \; \text{for all } g \in \mathcal{C}_{\text{TF},(\mathbf{i},\mathbf{j})}, m\in\mathcal{M}_{(\mathbf{i},\mathbf{j})},\\
    \text{[C3] }& \sum_{m\in\mathcal{M}_{(\mathbf{i},\mathbf{j})}}l_{g,m}\geq 1 \; \text{for all } g\in\mathcal{C}_{\text{TF},(\mathbf{i},\mathbf{j})}.
\end{split}
$}
\end{equation}
\noindent The decision variables are the binary variables $\{z_c\}$ and $\{l_{g,m}\}$, with $z_c = 1$ indicating combination $c$ is included in the cover and $z_c = 0$ otherwise. Here, the dummy objective of 1 is used, as we only wish to find a feasible solution satisfying constraints [C1]-[C3]. The first constraint [C1] requires the selected combinations $\{c: z_c = 1\}$ to \textit{cover} all failed test cases in $\mathcal{M}_{\ijbold}$. The next constraints [C2] and [C3] ensure the selected cover is indeed a \textit{minimal} cover. To see why, note that via constraint [C2], the auxiliary indicator variable $l_{g,m} \in \{0,1\}$ equals 1 if by removing $g$ from the considered cover, we fail to cover failure case $m$. For the considered cover to be minimal, we thus need, for each $g$ in the cover, at least one $l_{g,m} = 1$ for some failure case $m$; this is ensured by constraint [C3].


One appealing property of the integer feasible program \eqref{eqn:ILP} is that the objective is (trivially) linear and all constraints are linear in the binary decision variables. Such an integer \textit{linear} program thus admits nice structure for efficient large-scale optimization, particularly via recent developments in cutting plane and branch-and-bound algorithms \citep{balas1993lift,stidsen2014branch}. In our later implementation, we made use of the \texttt{GurobiPy} package in Python \citep{gurobi}, which implements state-of-the-art optimization solvers for large-scale integer programming. Gurobi is widely used for solving large-scale scheduling problems in the industry, including for the National Football League \citep{gurobi_NFL} and Air France \citep{gurobi_airfrance}. Here, with the ILP formulation \eqref{eqn:ILP}, Gurobi can solve for a feasible minimal set cover in seconds for our later case studies. This formulation thus provides an efficient strategy for computing the desired probability $\mathbb{P}(E_{(\mathbf{i},\mathbf{j})})$.

Of course, after finding a single minimal cover via \eqref{eqn:ILP}, we still have to find subsequent distinct minimal covers to compute \eqref{eq:cover}. This can easily be performed by iteratively solving the ILP \eqref{eqn:ILP} with an additional constraint that ensures subsequent covers are distinct from found covers. More concretely, let $\{\tilde{z}_c\}_{c \in \mathcal{C}_{\rm TF, \ijbold}}$ be a minimal cover found by \eqref{eqn:ILP}. Then a subsequent cover can be found by solving the ILP \eqref{eqn:ILP} with the additional constraint:
\begin{equation}
    d_c = z_c \oplus \tilde{z}_c, \quad  \sum_{c\in\mathcal{C}_{\text{TF},(\mathbf{i},\mathbf{j})}}d_c\geq 1, \quad c\in\mathcal{C}_{\text{TF},(\mathbf{i},\mathbf{j})},
    \label{eqn:ILP2}
    \tag*{[C4]}
\end{equation}
where $\oplus$ is the XOR operator. This new constraint [C4] ensures the next cover is distinct from the previous found cover. To see why, note that $d_c$ equals 1 only if the binary variables $z_c$ and $\tilde{z}_c$ are different; the inequality constraint in [C4] thus ensures all considered covers are different from the previous cover $\{\tilde{z}_c\}_{c \in \mathcal{C}_{\rm TF, \ijbold}}$. The resulting problem is still an integer linear program, as XOR can naturally be expressed as linear constraints \citep{magee1996integer}. More specifically, the XOR condition in [C4] can be equivalently expressed as:
\begin{equation}
    \sum_{c\in\mathcal{C}_{\text{TF},(\mathbf{i},\mathbf{j})}} (1-\tilde{z}_c) \circ z_c \ge 1,\quad  
    \sum_{c\in\mathcal{C}_{\text{TF},(\mathbf{i},\mathbf{j})}} \tilde{z}_c - \sum_{c\in\mathcal{C}_{\text{TF},(\mathbf{i},\mathbf{j})}} \tilde{z}_c \circ z_c \ge 1,
\end{equation}
where $\circ$ is the element-wise product. This
 is linear in the binary variables $\{z_c\}_{c \in \mathcal{C}_{\rm TF, \ijbold}}$.

With this, if a feasible solution is found for the ILP \eqref{eqn:ILP} with [C4], the optimization solver will return a distinct minimal cover, which we add to the collection. If not, the solver will instead return a ``dual certificate'' \citep{guzelsoy2010integer} that guarantees the ILP has no feasible solutions; such a certificate is possible via the linear nature of the integer program. One then iteratively solves the ILP \eqref{eqn:ILP} with constraint [C4] (modified to exclude two or more found covers) until the solver returns a dual certificate, in which case no feasible solutions are possible and thus all minimal covers have been enumerated.

\subsection{Computing Root Cause Probabilities}
After enumerating all minimal covers for $\mathcal{M}_{\ijbold}$, we can then compute $\mathbb{P}(E_{(\mathbf{i},\mathbf{j})})$ via Proposition \ref{prop:cover}. Let $\mathcal{V} = \{\tilde{\mathcal{C}}_1, \cdots, \tilde{\mathcal{C}}_{|\mathcal{V}|}\}$ be the collection of all minimal covers of $\mathcal{M}_{\ijbold}$ found by the above procedure. By the principle of inclusion-exclusion, it follows from \eqref{eq:cover} that:
\begin{equation}
\label{eq:incexc}
\resizebox{0.9\linewidth}{!}{%
$
\begin{split}
\mathbb{P}(E_{(\mathbf{i},\mathbf{j})}) &= \mathbb{P}(\{Z_c = 1 \textup{ for all } c \in \tilde{\mathcal{C}}\} \textup{, for at least one $\tilde{\mathcal{C}} \in \mathcal{V}$})\\
&= \sum_{\text{cover } \tilde{\mathcal{C}} \in \mathcal{V}} \; \prod_{c \in \tilde{\mathcal{C}}} p_c - \sum_{\text{covers } \tilde{\mathcal{C}},\tilde{\mathcal{C}}' \in \mathcal{V}} \; \prod_{c \in \tilde{\mathcal{C}} \cup \tilde{\mathcal{C}}'} p_c + \cdots + (-1)^{|\mathcal{V}|} \prod_{c \in {\tilde{\mathcal{C}}_1 \cup \cdots \cup \tilde{\mathcal{C}}_{|\mathcal{V}|}}}p_c,
\end{split}
$}
\end{equation}
where $p_c = \mathbb{P}(Z_c = 1)$ is again the prior root cause probability of combination $c$. We can then plug the computed $\mathbb{P}(E_{(\mathbf{i},\mathbf{j})})$ into Equation \eqref{eq:altform} to finally compute the desired root cause probability $\mathbb{P}(Z_{(\mathbf{i},\mathbf{j})}=1 | \mathcal{D})$ for a TF combination $(\mathbf{i},\mathbf{j})$.

For software systems with a small number of inputs, the set of minimal covers $\mathcal{V}$ may not be large, in which case the computation in \eqref{eq:incexc} would not be intensive. For larger systems with $|\mathcal{V}|$ large, one can employ the following second-order truncation as an approximation: 
\begin{equation}
\mathbb{P}(E_{(\mathbf{i},\mathbf{j})}) \approx \sum_{\text{cover } \tilde{\mathcal{C}} \in \mathcal{V}} \; \prod_{c \in \tilde{\mathcal{C}}} p_c - \sum_{\text{covers } \tilde{\mathcal{C}},\tilde{\mathcal{C}}' \in \mathcal{V}} \; \prod_{c \in \tilde{\mathcal{C}} \cup \tilde{\mathcal{C}}'} p_c,
\label{eq:incexc2}
\end{equation}
which bypasses the need for computing higher-order terms involving more than two covers. Note that, by the inclusion-exclusion principle, the right-hand side of \eqref{eq:incexc2} \textit{underestimates} the probability $\mathbb{P}(E_{(\mathbf{i},\mathbf{j})})$. This is by design: from \eqref{eq:altform}, this then results in a slight \textit{overestimation} of the posterior root cause probability $\mathbb{P}(Z_{(\mathbf{i},\mathbf{j})}=1 | \mathcal{D})$. From a risk perspective, this is preferable to an approximation procedure that underestimates such probabilities.

\subsection{Algorithm Summary}
For completeness, we provide in Algorithm \ref{alg:BayesFlo} a summary of the full BayesFLo procedure. Suppose a test set is performed, yielding outcome data $\mathcal{D} = \{(\mathbf{t}_m,y_m)\}_{m=1}^M$. Here, the test cases $\{\mathbf{t}_m\}_{m=1}^M$ should ideally be collected from a covering array to ensure good coverage of combinations, but this is not necessary for BayesFLo. With test data collected and priors elicited on the single-factor root cause probabilities $\{p_{(i,j)}\}_{i,j}$, we then partition the set of considered combinations $\mathcal{C}$ into TP, TF and UT combinations using Equation \eqref{eq:classify}. Here, if the test engineer is confident that a root cause should not exceed a certain order, then posterior probabilities need only to be computed for combinations up to such an order. This can be justified as a stronger form of combination hierarchy, and can further reduce computation for evaluating posterior probabilities. 

Next, we compute posterior root cause probabilities within each category. For TP combinations, this is trivially zero as such combinations were cleared in passed cases. For UT combinations, this can be set as the prior probabilities from \eqref{eqn:prior_propagation}, as no information can be gleaned on such combinations from the test data. For TF combinations, their posterior probabilities can be computed via the minimal set cover approach in Section \ref{sec:posterior}. Finally, we can then rank the potential root causes in terms of their posterior probabilities, which can be used for planning downstream debugging efforts; more on this later.


{While it is difficult to establish the theoretical complexity of our algorithm, our implementation in later case studies suggests that it is quite efficient in practice. For both case studies (the first with $I=12$ factors and two failed test runs, and the second with $I=15$ factors and six failed test runs), the desired BayesFLo posterior probabilities can be computed within a minute on a single-core desktop processor. This shows that, with a careful integration of recent integer programming and graph representation tools, one can efficiently perform Bayesian inference for this challenging high-dimensional fault localization problem. With the availability of multi-core processors, BayesFLo can be further accelerated by exploiting such parallelism within the \texttt{GurobiPy} package; see \cite{glockner2015parallel} for details.}

Finally, we discuss how the posterior probabilities from BayesFLo can help guide actionable decisions for \textit{confident} fault localization. One important use is for planning the subsequent debugging effort for test engineers. Let $\alpha$ be a user-specified significance level, where suspicious combinations with posterior probabilities below $\alpha$ can be confidently ignored for further investigation. A default choice may be $\alpha = 0.05$; a smaller choice can be used if the tester is more risk-averse. With this, one can then plan the required debugging effort by testing all combinations with probabilities greater than $\alpha$ (call this set $\mathcal{S}$), beginning with the ones with highest probabilities. When $\mathcal{S}$ is small, this suggests just a small debugging effort can confidently identify the root cause. When $\mathcal{S}$ is large, this suggests a massive debugging effort may be needed, which indicates the software is not ready for release in the near future. Such insights can help test engineers forecast the required resources and timeframe for reliable software development.

\begin{algorithm}[!t]
\caption{BayesFLo: Bayesian Fault Localization}\label{alg:BayesFlo}
\begin{justify}
\vspace{-0.2cm}
\textbf{Input:} Test data $\mathcal{D} = \{(\mathbf{t}_m,y_m)\}_{m=1}^M$, consisting of each test case and its corresponding test outcome.\\
\textbf{Output:} Potential root causes $(\mathbf{i},\mathbf{j}) \in \mathcal{C}$ with posterior root cause probabilities $\mathbb{P}(Z_{(\mathbf{i},\mathbf{j})}=1|\mathcal{D})$.
\end{justify}
\vspace{-0.2cm}
\begin{algorithmic}[1]
    \State Elicit root cause probabilities $\{p_{(i,j)}\}_{i,j}$ from domain knowledge.
    \State Partition the set of considered combinations $\mathcal{C}$ into TP, TF and UT combinations using Equation \eqref{eq:classify}.
    \State For TP combinations, set its posterior root cause probability to 0.
    \State For UT combinations, set its posterior root cause probability as the prior probability \eqref{eqn:prior_propagation}.
    \State For each TF combination $(\mathbf{i},\mathbf{j})$, enumerate minimal covers for the failed cases in $\mathcal{M}_{(\mathbf{i},\mathbf{j})}$, then compute its posterior root cause probability using Equations \eqref{eq:altform} and \eqref{eq:incexc2}.
    \State Rank potential root causes (TF and UT combinations) using its corresponding posterior probabilities.
\end{algorithmic}
\end{algorithm}

\section{Case Studies}
\label{sec:case_studies}

We now investigate the effectiveness of BayesFLo in two case studies: the first on our motivating application on the Traffic Alert and Collision Avoidance System, and the second on the Vulnerable Road User protection tests for safe autonomous driving.




\subsection{Case Study 1: Traffic Alert and Collision Avoidance System}
\label{sec:TCAS}


Consider first the motivating case study in Section \ref{sec:mot}, on the fault localization of the Traffic Alert and Collision Avoidance System (TCAS). TCAS targets the avoidance of potential mid-air collisions between aircraft, and its reliable performance is essential for mitigating catastrophic disasters (see, e.g., \citealp{dccrash}). In what follows, we consider a specific program -- the Siemens test suite in \cite{griesmayer2007automated} -- for fault localization. Its $I=12$ input factors with corresponding levels are summarized in Table \ref{tab:TCAS_vars}. We adopt the ``TCAS v2'' set-up from \cite{griesmayer2007automated}, which contains a fully-documented fault that results in an incorrect resolution advisory; if unaddressed, such a fault may lead to potential aircraft collisions. Here, it is known \citep{griesmayer2007automated} that such a failure arises from a single root cause, namely, the 3-FC involving factors 12, 8 and 9:
\begin{equation}
\small
(\texttt{Climb\_Inhibit} = 1)\ \times\ (\texttt{Up\_Separation} = 399)\ \times\ (\texttt{Down\_Separation} = 640). 
    \label{eq:TCAS_RC}
\end{equation}
\normalsize
This failure was found to arise as follows. 
Imagine two aircraft (one primary, the other incoming) are headed towards each other. Suppose the incoming aircraft is currently below but is projected to be above the primary aircraft at the closest approach \citep{bradley1992simulation}. When the climb rate is restricted for the primary aircraft (i.e., $\texttt{Climb\_Inhibit}=1$), there is a limit to the height that it can climb. In TCAS, a constant is thus added to the projected vertical separation between aircraft at closest approach due to an upward maneuver of the primary aircraft (i.e., \texttt{Up\_Separation}), to ensure a resolution advisory is raised if it exceeds a projected downward maneuver (i.e., \texttt{Down\_Separation}). Here, an incorrect constant was added in the system, resulting in resolution advisories not being raised when they should have been, thus precipitating a failure.
We will investigate the effectiveness of BayesFLo compared to the state-of-the-art in identifying the root cause \eqref{eq:TCAS_RC} with limited test runs.

With this set-up, we then perform test runs using the strength-2 covering array from Table \ref{tab:TCAS_vars}, with a run size of $M=19$. Table \ref{tab:TCAS_design} shows 
the corresponding test outcomes for each test run. Table \ref{tab:TCAS_RefRes} summarizes the ranked suspicious combinations using the BEN approach \citep{ghandehari2018combinatorial} and the weighting approach \citep{lekivetz2018fault}, which serve as the state-of-the-art for combinatorial fault localization in our experiments. Recall the two limitations noted in Section \ref{sec:mot}. First, both approaches need to investigate eight suspicious combinations before identifying the true root cause, because they both require the inspection of 2-factor combinations before 3-factor ones. This can be undesirable given the costly nature of debugging each combination \citep{landsberg2018doric}. Second, while both approaches provide a ranked list of suspicious combinations, the employed metrics cannot be interpreted as probabilities conditional on test data, which may hinder actionable decisions for confident fault localization. We explore next how BayesFLo can tackle these two limitations.



To implement BayesFLo, we need a careful specification of its prior distribution in \eqref{eqn:prior_propagation}. Here, prior domain knowledge on the nature of the incorrect advisory fault (described earlier) suggests that it is likely associated with the projected vertical separation between two aircraft, which involves the two ``separation'' factors \texttt{Up\_Separation} and \texttt{Down\_Separation}. To reflect this, we adopt the following informative prior. We set $p_{(i,j)}=1/30$ as the baseline prior probability, where its denominator $\sum_{i=1}^IJ_i=30$ is the total number of levels for all $I=12$ factors. This baseline prior is assigned to all factors except for \texttt{Up\_Separation} and \texttt{Down\_Separation}; the latter is assigned a prior of $p_{(i,j)}=2/30$ to reflect heightened prior suspicion. We will contrast this with an ``uninformative'' implementation of BayesFLo, where all priors are assigned the same prior probability $p_{(i,j)}=1/30$. 
Posterior root cause probabilities are then computed via Algorithm \ref{alg:BayesFlo}. For the weighting approach in \cite{lekivetz2018fault}, the weights for the two ``separation'' factors are similarly set to be double that for the other factors to reflect this prior domain knowledge.


\begin{figure}[!t]
  \centering
  \begin{subfigure}[t]{0.48\textwidth}
    \centering
    \scalebox{0.85}{
    \begin{tabular}{ccc}
      \toprule
      \textbf{Combination Type} & \textbf{Count} & \textbf{Probability}\\
      \midrule
      \textbf{3-FC} & \textbf{1} & \textbf{0.55}\\
      2-FC & 1 & 0.41\\
      2-FC & 7 & 0.10\\
      3-FC & 140 & $<$0.01\\
      \bottomrule
    \end{tabular}}
    \caption{BayesFLo (uninformative prior).}
  \end{subfigure}\hfill
  \begin{subfigure}[t]{0.48\textwidth}
    \centering
    \scalebox{0.85}{
   \begin{tabular}{ccc}
      \toprule
      \textbf{Combination Type} & \textbf{Count} & \textbf{Probability} \\
      \midrule
      \textbf{3-FC} & \textbf{1} & \textbf{0.65}\\
      2-FC & 1 & 0.42\\
      2-FC & 2 & 0.14\\
      2-FC & 5 & 0.07\\
      3-FC & 140 & $<$0.03\\
      \bottomrule
    \end{tabular}}
    \caption{BayesFLo (informative prior).}
  \end{subfigure}
  \captionof{table}{Suspicious 2-FCs and 3-FCs ranked by their posterior root cause probabilities using BayesFLo (with uninformative and informative priors) for the TCAS case study. Combinations at the top are more suspicious and are investigated first. The true root cause is \textbf{bolded}.}
  \label{tab:TCAS_BayesFLo}
\end{figure}

Table \ref{tab:TCAS_BayesFLo} shows the 2-factor and 3-factor suspicious combinations from BayesFLo (using the above informative and uninformative prior specifications), ranked by their posterior root cause probabilities. Here, for all compared methods, we consider only suspicious combinations with at most three factors; this not only reduces computation, but is also guided by prior knowledge that root causes are typically of lower order for such systems \citep{kuhn2004software}. Out of 149 suspicious 2-FCs and 3-FCs, BayesFLo (with an informative prior) is able to pinpoint the true 3-factor root cause \eqref{eq:TCAS_RC} as its most suspicious combination, with a posterior probability of 0.65. Two 2-FCs are ranked next with probabilities of 0.42 and 0.14, the remaining 2-FCs are ranked next with probabilities 0.07, and the remaining 3-FCs are ranked last with probabilities less than 0.03. From a debugging perspective, the BayesFLo analysis is preferable to existing techniques: it finds the true root cause as its top-ranked combination, without needing to further investigate and rule out the eight 2-FCs via expensive tests. This shows the importance of a Bayesian statistical framework that captures domain and structural knowledge, allowing for a principled comparison of suspicious combinations across different orders for efficient fault localization. Comparing BayesFLo with the informative vs. uninformative prior, we see that the incorporation of domain knowledge (i.e., the heightened suspiciousness of the two separation factors) raises the posterior probability of the true root cause, which shows that incorporating such knowledge can indeed enhance fault localization.

It is also important to investigate the probabilistic aspect of the BayesFLo analysis, and how this can help guide actionable decisions for confident fault localization. Recall that for existing analyses (see Table \ref{tab:TCAS_RefRes}), which do not employ probabilistic metrics, it may be difficult to plan the subsequent debugging effort, since it is not clear which 2-FCs or 3-FCs can be confidently ignored for debugging. Here, BayesFLo can give a clearer picture. Using the recommended significance level $\alpha = 0.05$, the (informative) BayesFLo analysis suggests only nine combinations need to be tested; the last 140 3-FCs can be confidently ignored, since their posterior probabilities are less than $\alpha$. This is quite reasonable given test outcomes, where only two failed test runs were observed. The probabilistic analysis from BayesFLo thus enables effective planning of debugging efforts for confident fault localization.

{\subsection{Case Study 2: Vulnerable Road User Protection Tests}
\label{sec:vru}


{Consider next a case study on the Vulnerable Road User protection test for safe autonomous driving. This case study is adapted from \cite{Gladisch2020}, where the goal is to test and validate safety standards for automated driving systems via the Vulnerable Road User protection tests. VRU tests evaluate how effective the autonomous emergency braking system is at detecting and reacting to pedestrians and cyclists under a broad range of driving conditions. The VRU scenario considered in \cite{Gladisch2020} involves a child jaywalking across an intersection in front of an approaching vehicle. Here, we consider a system with $I=15$ factors, listed in Table \ref{tab:VRU_vars} with their corresponding levels. The first ten factors were investigated in \cite{Gladisch2020}, and the last five are new factors that we added in our case study. Unfortunately, \cite{Gladisch2020} explores the problem of test case design and does not shed light on specific root causes. As a proof-of-concept, we consider the following 2-FC as the root cause:
\begin{equation}
  (\texttt{Reflection on road} = \texttt{yes})\ \times\ (\texttt{Speed} = \texttt{fast}).
  \label{eq:VRU_rc}
\end{equation}
This is quite intuitive: under fast driving speeds and high reflection from the road surface, the automatic driving system can fail to detect the jaywalking child, resulting in a failure.

\begin{figure}[!t]
    \centering
    \scalebox{0.9}{
    \begin{tabular}{c c c }
        \toprule
        \textbf{Factor} & \textbf{Description} & \textbf{Levels} \\
        \hline
        1 & \texttt{Daytime} & morning, day, evening, night\\
        2 & \texttt{Haze/Fog} & no, yes\\
        3 & \texttt{Street condition} & dry, wet, icy, snow, broken\\
        4 & \texttt{Sky} & cloudy, no, clear\\
        5 & \texttt{Rain} & no, yes\\
        6 & \texttt{Reflection on road} & no, yes\\
        7 & \texttt{Shadow on road} & no, yes\\
        8 & \texttt{VRU type} & adult, child\\
        9 & \texttt{VRU pose} & pedestrian, jogger, cyclist\\
        10 & \texttt{VRU contrast to BG} & low, high\\
        11 & \texttt{Speed} & slow, fast\\
        12 & \texttt{Vehicle size} & small, medium, large\\
        13 & \texttt{Lane number} & two, four\\
        14 & \texttt{Camera} & no, yes\\
        15 & \texttt{Direction} & same, opposite\\
        \toprule
    \end{tabular}}
    \captionof{table}{Input factors and their corresponding levels for the VRU case study.}
    \label{tab:VRU_vars}
\end{figure}



With this, we perform test runs using a strength-2 covering array with $M=23$ runs. These runs and their test outcomes are provided in Appendix C for brevity, where six runs resulted in failures. The BEN approach in \cite{ghandehari2018combinatorial} is then applied to the test outcomes, with its top-ranked suspicious combinations summarized in Table \ref{tab:VRU_BayesFLo}(a). Implementing the weighting approach in \cite{lekivetz2018fault}, however, comes with computational challenges. Recall from Section \ref{sec:mot} that the computation of its normalized weights requires a complete enumeration of potential root cause scenarios, which can be infeasible when there are many failed test runs. Here, the six failed runs in our test set (see Appendix C) give rise to 54 2-factor potential root causes; a complete enumeration involves $2^{54}=1.80\times 10^{16}$ possible scenarios, which is infeasible to perform. We thus do not compare with the weighting approach benchmark in this case study.

As before, our BayesFLo approach is implemented under an informative and uninformative prior specification. For the informative prior, we adopt the baseline prior probability $p_{(i,j)}=1/38$, where its denominator $\sum_{i=1}^I J_i = 38$ is the total number of levels over all $I$ factors. This baseline prior is assigned to the first ten factors (from the original system in \citealp{Gladisch2020}), and a higher prior $p_{(i,j)}=2/38$ is assigned to the last five factors, which are newly added and have not been thoroughly tested. For the uninformative prior, all 15 factors are assigned the baseline prior $p_{(i,j)}=1/38$.

Table \ref{tab:VRU_BayesFLo} (b) and (c) show the suspicious 2-FCs from the informative and uninformative BayesFLo approaches, ranked by their posterior root cause probabilities. Here, all 3-FCs have lower probabilities than the lowest-ranked 2-FC, and are thus not reported for brevity. We see that the informative BayesFLo approach is able to pinpoint the true 2-factor root cause \eqref{eq:VRU_rc} as its most suspicious combination, with a posterior probability of 0.67. The next four 2-FCs have posterior probabilities between 0.28 and 0.44, and the remaining 2-FCs have probabilities less than 0.25. As before, BayesFLo is preferable to the BEN approach (see Table \ref{tab:VRU_BayesFLo}(a)) from a debugging perspective: it finds the true root cause as its top combination, whereas BEN requires the investigation of four additional combinations before finding the root cause. There are two reasons for this: (i) BEN does not incorporate prior domain knowledge in its analysis, and (ii) BEN relies on single-factor components for evaluating its suspiciousness metrics (see Section \ref{sec:mot}), which may obscure the identification of multi-factor root causes. Comparing the informative BayesFLo with its uninformative counterpart, we again see that the incorporation of prior domain knowledge can improve fault localization: such knowledge helps break the tie of the two top 2-FCs in the uninformative analysis, allowing BayesFLo to pinpoint the true root cause as its top combination.

Finally, this probabilistic BayesFLo analysis also helps guide actionable decisions for confident fault localization. From the BEN analysis (see Table \ref{tab:VRU_BayesFLo}(a)), it is again difficult to plan the subsequent debugging effort, since it is not clear which combinations can be confidently ignored. BayesFLo can give a clearer picture. With a significance level of $\alpha = 0.05$, the informative BayesFLo analysis suggests that 39 combinations need to be further investigated; the last 15 2-FCs and all 3-FCs (with probabilities less than $\alpha$) can be confidently ignored. Note that the set of combinations to debug here is larger than that from the earlier TCAS case study. This is quite intuitive: there are considerably more failed test runs here, which should result in heightened suspicion and a larger debugging effort. Our probabilistic BayesFLo analysis thus captures this intuition to guide an effective planning of debugging efforts.


}

\begin{figure}[!t]
  \centering
  \begin{subfigure}[t]{0.32\textwidth}
    \centering
    \scalebox{0.85}{
   \begin{tabular}{cccc}
      \toprule
      \textbf{Type} & \textbf{Count} & \textbf{Rank}\\
      \midrule
      2-FC & 4 & 1-4\\
      \textbf{2-FC} & \textbf{1} & \textbf{5}\\
      2-FC & 49 & 6-54\\
      \bottomrule
    \end{tabular}}
    \caption{BEN Phase 1.}
  \end{subfigure}
  \begin{subfigure}[t]{0.32\textwidth}
    \centering
    \scalebox{0.85}{
    \begin{tabular}{cccc}
      \toprule
      \textbf{Type} & \textbf{Count} & \textbf{Probability}\\
      \midrule
      \textbf{2-FC} & \textbf{1} & \textbf{0.50}\\
      2-FC & 1 & 0.50\\
      2-FC & 2 & 0.33\\
      2-FC & 50 & 0.05-0.25\\
      \bottomrule
    \end{tabular}}
    \caption{BayesFLo (uninformative).}
  \end{subfigure}\hfill
  \begin{subfigure}[t]{0.32\textwidth}
    \centering
    \scalebox{0.85}{
   \begin{tabular}{cccc}
      \toprule
      \textbf{Type} & \textbf{Count} & \textbf{Probability}\\
      \midrule
      \textbf{2-FC} & \textbf{1} & \textbf{0.67}\\
      2-FC & 1 & 0.44\\
      2-FC & 1 & 0.33\\
      2-FC & 1 & 0.32\\
      2-FC & 2 & 0.28\\
      2-FC & 33 & 0.05-0.25\\
      2-FC & 15 & $<$0.05\\
      \bottomrule
    \end{tabular}}
    \caption{BayesFLo (informative).}
  \end{subfigure}
  \captionof{table}{Suspicious 2-FCs ranked using the BEN approach (Phase 1) and BayesFLo (with uninformative and informative priors) for the VRU case study. Combinations at the top are deemed more suspicious and are investigated first. The true root cause is \textbf{bolded}.}
  \label{tab:VRU_BayesFLo}
\end{figure}

}

\section{Conclusion}
\label{sec:conclusion}


We proposed a new BayesFLo framework for Bayesian fault localization of complex software systems. Compared to existing methods, BayesFLo can accelerate the identification of root causes with a small number of test runs, and provide a principled quantification of probabilistic risk that can guide subsequent debugging efforts. This is achieved via a new Bayesian fault localization model, which embeds the combination hierarchy and heredity principles \citep{lekivetz2021testing} to capture the structured nature of software root causes. One challenge is the computation of posterior root cause probabilities for BayesFLo, which can be infeasible even for small systems. We thus developed a new algorithmic framework for computing the desired posterior probabilities, leveraging recent tools from integer programming and graph representations. We then demonstrate the effectiveness of BayesFLo over the state-of-the-art in case studies on the Traffic Alert Collision Avoidance System and the Vulnerable Road User protection tests.


Building on these promising results, there are many immediate avenues for future work. One direction is the use of the BayesFLo framework for sequential design of subsequent test sets. This adaptive testing of software, facilitated by the proposed Bayesian model, can accelerate the discovery of bugs in complex systems; recent work in \cite{chen2022adaptive} appears to be useful for this goal. Another direction is the extension of BayesFLo for fault localization of systems with mixed (i.e., continuous and discrete) factors. Such a setting would be more complex, as it requires the probabilistic modeling of the fault response surface. Finally, in the rare scenario where one encounters multiple failure types, a reasonable approach may be a Bayesian fault localization framework with categorical responses that can pinpoint root causes for different failure types; we will explore this in future work.\\


\noindent \textbf{Acknowledgements.} This work was supported by NSF CSSI 2004571, NSF DMS 2210729, NSF DMS 2220496, NSF DMS 2316012 and DE-SC0024477.


\newpage
\spacingset{1.0}
\bibliography{references}

\newpage
\spacingset{1.45}
\begin{appendix}
\section{Proof for Proposition 1}
\noindent \textit{Proof}: We first define the following two index sets for failed test cases:
\begin{align*}
\begin{split}
    \mathcal{M}_{(\mathbf{i},\mathbf{j})} &= \{m = 1, \cdots, M : y_m = 1, (\mathbf{i},\mathbf{j}) \in \mathcal{C}_{{\rm TF},m}\},\\
    \mathcal{M}_{-(\mathbf{i},\mathbf{j})} &= \{m = 1, \cdots, M : y_{m} = 1, (\mathbf{i},\mathbf{j}) \notin \mathcal{C}_{{\rm TF},m}\}.\\
\end{split}
\end{align*}
For a given tested-and-failed (TF) combination $(\mathbf{i},\mathbf{j})$, $\mathcal{M}_{(\mathbf{i},\mathbf{j})}$ and $\mathcal{M}_{-(\mathbf{i},\mathbf{j})}$ split the failed test cases into cases that either involve or do not involve $(\mathbf{i},\mathbf{j})$.

Next, we define the following events:
\small
\begin{align*}
\begin{split}
E_{\text{P}} &= \{ \text{$Z_c = 0$ for all $c \in \mathcal{C}_{\rm TP}$}\},\\
E_{(\mathbf{i},\mathbf{j})} &= \{\textup{for each $m \in \mathcal{M}_{(\mathbf{i},\mathbf{j})}$, there exists some $c \in \mathcal{C}_{{\rm TF},m} \setminus \mathcal{C}_{\rm TP} $ such that $Z_c = 1$} \},\\
E_{-(\mathbf{i},\mathbf{j})} &= \{\textup{for each $m' \in \mathcal{M}_{-(\mathbf{i},\mathbf{j})}$, there exists some $c \in \mathcal{C}_{{\rm TF},m'} \setminus \mathcal{C}_{\rm TP} $ such that $Z_c = 1$} \}.
\end{split}
\end{align*}
\normalsize
Here, $E_{(\mathbf{i},\mathbf{j})}$ is the event that all failures in $\mathcal{M}_{(\mathbf{i},\mathbf{j})}$ can be explained by some TF combination as a root cause, and $E_{-(\mathbf{i},\mathbf{j})}$ is the event that all failures in $\mathcal{M}_{-(\mathbf{i},\mathbf{j})}$ can be explained by some TF combination as a root cause. Thus, the observed test data $\mathcal{D}$ is equivalent to the intersection of the events $E_{\text{P}}$, $E_{(\mathbf{i},\mathbf{j})}$ and $E_{-(\mathbf{i},\mathbf{j})}$.

As $(\mathbf{i},\mathbf{j})$ is not contained in $E_{\text{P}}$ and $E_{-(\mathbf{i},\mathbf{j})}$ by construction, it follows that  $Z_{(\mathbf{i},\mathbf{j})}\perp E_{\text{P}}$ and $Z_{(\mathbf{i},\mathbf{j})}\perp E_{-(\mathbf{i},\mathbf{j})}$. As such, the desired posterior root cause probability can be simplified as:
\begin{align*}
    \mathbb{P}(Z_{(\mathbf{i},\mathbf{j})}=1 | \mathcal{D})  
    & = \mathbb{P}(Z_{(\mathbf{i},\mathbf{j})}=1 | E_{(\mathbf{i},\mathbf{j})}).
    \label{eqn:probE}
\end{align*}

\noindent Next, note that $\mathbb{P}(E_{\ijbold}|Z_{\ijbold}=1)=1$ as $\ijbold$ is contained in every failure case $m\in\mathcal{M}_{\ijbold}$. We thus get:
\begin{align*}
\begin{split}
    \mathbb{P}(Z_{(\mathbf{i},\mathbf{j})}=1 | E_{(\mathbf{i},\mathbf{j})}) &= \frac{\mathbb{P}(Z_{\ijbold}=1)\cdot 1}{\mathbb{P}(E_{(\mathbf{i},\mathbf{j})})}\\
    &= \frac{p_{(\mathbf{i},\mathbf{j})}}{\mathbb{P}(E_{(\mathbf{i},\mathbf{j})})},
    \end{split}
\end{align*}
which is as desired.
\qed

\newpage
\section{Proof for Proposition 2}
\noindent \textit{Proof}: Recall that $E_{(\mathbf{i},\mathbf{j})}$ is defined as:
\[E_{(\mathbf{i},\mathbf{j})}=\{\textup{for each $m \in \mathcal{M}_{(\mathbf{i},\mathbf{j})}$, there exists some $c \in \mathcal{C}_{{\rm TF},m} \setminus \mathcal{C}_{{\rm TP}}  $ such that $Z_c = 1$} \}.
\]
In words, this is the event that each failed test case in  $\mathcal{M}_{(\mathbf{i},\mathbf{j})}$ can be explained by some TF combination as a root cause. Define $F_{(\mathbf{i},\mathbf{j})}$ as the event:
\[F_{(\mathbf{i},\mathbf{j})}= \{\{Z_c = 1 \textup{ for all } c \in \tilde{\mathcal{C}}\} \textup{, for at least one minimal cover $\tilde{\mathcal{C}}$ of $\mathcal{M}_{(\mathbf{i},\mathbf{j})}$}\}.\]
We wish to show that $E_{(\mathbf{i},\mathbf{j})} = F_{(\mathbf{i},\mathbf{j})}$.

Consider first $E_{(\mathbf{i},\mathbf{j})} \subseteq F_{(\mathbf{i},\mathbf{j})}$. This must be true, since if we collect all TF combinations $c$ with $Z_c=1$ from $E_{\ijbold}$, they contain at least one minimal cover $\tilde{\mathcal{C}}$, as all failures $m\in\mathcal{M}_{\ijbold}$ are covered. This suggests that every arbitrary element in $E_{\ijbold}$ is also an element in $F_{\ijbold}$. Consider next $F_{(\mathbf{i},\mathbf{j})} \subseteq E_{(\mathbf{i},\mathbf{j})}$. This must also be true, since if we take a minimal cover $\tilde{\mathcal{C}}$ of $\mathcal{M}_{\ijbold}$ with $Z_c=1$ for all $c\in\tilde{\mathcal{C}}$ from $F_{\ijbold}$, then for every $m\in\mathcal{M}_{\ijbold}$ there exists at least one $c\in\tilde{\mathcal{C}}$ that explains this failure. Thus every arbitrary element in $F_{\ijbold}$ is also an element in $E_{\ijbold}$. This proves the proposition.
\qed

\newpage
\section{Test design and outcomes for the VRU case study}
\begin{figure}[!h]
    \centering
    \scalebox{0.9}{
    \begin{tabular}{c ccccccccccccccc c}
        \toprule
        & \multicolumn{15}{c}{\textbf{Factor}} &  \\
        \textbf{Run Number} & \textbf{1} & \textbf{2} & \textbf{3} & \textbf{4} & \textbf{5} & \textbf{6} & \textbf{7} & \textbf{8} & \textbf{9} & \textbf{10} & \textbf{11} & \textbf{12} & \textbf{13} & \textbf{14} & \textbf{15} & \textbf{Outcome}\\
        \hline
        1 & 4 &	1 &	1 &	3 &	2 &	2 &	1 &	1 &	3 &	1 &	1 &	3 &	2 &	2 &	2 & 0\\
        2 & 4 &	1 &	2 &	2 &	2 &	2 &	2 &	1 &	2 &	2 &	2 &	2 &	2 &	1 &	2 & 1\\
        3 & 3 &	2 &	3 &	3 &	1 &	2 &	1 &	1 &	2 &	2 &	1 &	3 &	1 &	1 &	2 & 0\\
        4 & 1 &	1 &	1 &	1 &	1 &	2 &	1 &	2 &	1 &	2 &	1 &	1 &	1 &	1 &	2 & 0\\
        5 & 2 &	2 &	5 &	3 &	1 &	1 &	2 &	1 &	2 &	2 &	2 &	2 &	2 &	1 &	2 & 0\\
        6 & 4 &	2 &	4 &	3 &	1 &	2 &	1 &	2 &	3 &	2 &	1 &	1 &	2 &	2 &	1 & 0\\
        7 & 3 &	2 &	5 &	2 &	2 &	2 &	1 &	1 &	3 &	1 &	2 &	3 &	2 &	1 &	1 & 1\\
        8 & 2 &	2 &	1 &	2 &	1 &	2 &	2 &	2 &	2 &	1 &	1 &	2 &	1 &	1 &	1 & 0\\
        9 & 4 &	1 &	3 &	3 &	1 &	1 &	2 &	2 &	1 &	1 &	2 &	2 &	1 &	2 &	1 & 0\\
        10 & 1 & 1 & 5 & 3 & 2 & 1 & 1 & 2 & 2 & 1 & 1 & 1 & 2 & 2 & 1 & 0\\
        11 & 3 & 2 & 1 & 1 & 1 & 1 & 2 & 1 & 1 & 1 & 2 & 2 & 2 & 2 & 1 & 0\\
        12 & 4 & 2 & 5 & 1 & 2 & 1 & 2 & 2 & 1 & 2 & 1 & 3 & 1 & 1 & 1 & 0\\
        13 & 2 & 1 & 3 & 1 & 1 & 1 & 2 & 1 & 2 & 2 & 2 & 1 & 1 & 2 & 1 & 0\\
        14 & 1 & 2 & 4 & 2 & 2 & 2 & 1 & 1 & 3 & 1 & 2 & 2 & 2 & 1 & 2 & 1\\
        15 & 2 & 1 & 4 & 2 & 2 & 2 & 2 & 2 & 1 & 1 & 2 & 3 & 1 & 1 & 2 & 1\\
        16 & 2 & 2 & 2 & 1 & 1 & 2 & 1 & 2 & 3 & 1 & 1 & 1 & 1 & 2 & 1 & 0\\
        17 & 1 & 1 & 2 & 2 & 2 & 2 & 2 & 2 & 1 & 2 & 2 & 3 & 2 & 2 & 2 & 1\\
        18 & 1 & 1 & 3 & 2 & 1 & 1 & 2 & 2 & 3 & 1 & 2 & 1 & 2 & 2 & 2 & 0\\
        19 & 3 & 1 & 4 & 3 & 1 & 1 & 1 & 2 & 2 & 1 & 2 & 1 & 1 & 2 & 2 & 0\\
        20 & 3 & 1 & 2 & 2 & 2 & 2 & 1 & 2 & 3 & 1 & 2 & 1 & 2 & 2 & 1 & 1\\
        21 & 2 & 2 & 2 & 3 & 1 & 1 & 1 & 2 & 3 & 1 & 1 & 1 & 1 & 2 & 1 & 0\\
        22 & 4 & 2 & 4 & 1 & 1 & 2 & 1 & 2 & 3 & 2 & 1 & 1 & 2 & 2 & 1 & 0\\
        23 & 2 & 1 & 3 & 1 & 2 & 1 & 2 & 1 & 2 & 2 & 2 & 1 & 1 & 2 & 1 & 0\\
        \toprule
    \end{tabular}}
    \captionof{table}{The $M=23$-run CA design and test outcomes for the VRU case study. Here, an outcome of 0 indicates a passed test case, with 1 indicating a failed test case.}
    \label{tab:VRU_design}
\end{figure}

\end{appendix}
\end{document}